\documentclass[final,journal]{IEEEtran}

\usepackage[boxed,linesnumbered]{algorithm2e}
\usepackage{amsmath}
\usepackage{amssymb}
\usepackage{balance}
\usepackage{bm}
\usepackage{cite}
\usepackage{color}
\usepackage{enumitem}
\usepackage[T1]{fontenc}
\usepackage{gensymb}
\usepackage{graphicx}
\usepackage[utf8]{inputenc}
\usepackage{longtable}
\usepackage{mathrsfs}
\usepackage{mathtools}
\usepackage{multirow}
\usepackage[cmintegrals]{newtxmath}
\usepackage{siunitx}
\usepackage{soul}
\usepackage{tabularx}
\usepackage[color=green!40]{todonotes}
\soulregister\cite7
\soulregister\footnote7
\soulregister\eqref7
\soulregister\ref7
\usepackage{threeparttable}
\usepackage{url}
\usepackage{hyperref}
\hypersetup{hidelinks}
\newlength{\myvspace}

\graphicspath{{./}{../figures/}}

\newcommand{\orcidA}[1]{\href{https://orcid.org/0000-0002-8382-785X}{\includegraphics[width=10pt]{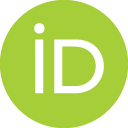}}}
\newcommand{\orcidB}[1]{\href{https://orcid.org/0000-0003-4231-7455}{\includegraphics[width=10pt]{ORCIDiD_icon.png}}}
\newcommand{\orcidC}[1]{\href{https://orcid.org/0000-0002-4123-2647}{\includegraphics[width=10pt]{ORCIDiD_icon.png}}}
\newcommand{\orcidD}[1]{\href{https://orcid.org/0000-0002-0212-2365}{\includegraphics[width=10pt]{ORCIDiD_icon.png}}}
\newcommand{\myauthor}{%
  Zhe Tang\orcidA{0000-0002-8382-785X},~\IEEEmembership{Student~Member,~IEEE},%
  Sihao Li\orcidB{0000-0003-4231-7455},~\IEEEmembership{Student~Member,~IEEE},%
  Kyeong Soo
  Kim\orcidC{0000-0002-4123-2647},~\IEEEmembership{Senior~Member,~IEEE}, and%
  Jeremy
  Smith\orcidD{0000-0002-0212-2365},~\IEEEmembership{Senior~Member,~IEEE}%
}%
\newcommand{\mytitle}{On the Multidimensional Augmentation of Fingerprint Data
  for Indoor Localization in A Large-Scale Building Complex Based on
  Multi-Output Gaussian Process}%

\begin{document}

\title{\mytitle}

\author{%
  \myauthor%
  \thanks{%
    An earlier version of this paper was presented in part at the IEEE Fourth
    International Workshop on Data Driven Intelligence for Networks and Systems
    (DDINS) organized in conjunction with IEEE International Conference on
    Communications (ICC) 2022, Seoul, Korea, May 2022.%
  }%
  \thanks{%
    Z.~Tang and S.~Li are with the Department of Electrical Engineering and
    Electronics, University of Liverpool, Liverpool L69 3GJ, U.K., and also with
    the School of Advanced Technology, Xi'an Jiaotong-Liverpool University,
    Suzhou 215123, P. R. China (e-mail: Zhe.Tang@liverpool.ac.uk;
    Sihao.Li@liverpool.ac.uk).

    K.~S.~Kim is with the School of Advanced Technology, Xi'an Jiaotong-Liverpool
    University, Suzhou 215123, P. R. China (e-mail: Kyeongsoo.Kim@xjtlu.edu.cn).

    J.~S.~Smith is with the Department of Electrical Engineering and Electronics,
    University of Liverpool, Liverpool L69 3GJ, U.K. (e-mail:
    J.S.Smith@liverpool.ac.uk). } }%%%

% \markboth{IEEE Transactions on Network and Service Management, Vol.~20, Issue~1,
%   March 2023}{Tang \MakeLowercase{et al.}: \mytitleshort}

\maketitle

\begin{abstract}
  Wi-Fi fingerprinting becomes a dominant solution for large-scale indoor
  localization due to its major advantage of not requiring new infrastructure
  and dedicated devices. The number and the distribution of Reference Points
  (RPs) for the measurement of localization fingerprints like Received Signal
  Strength Indicator (RSSI) during the offline phase, however, greatly affects
  the localization accuracy; for instance, the UJIIndoorLoc---i.e., the
  publicly-available multi-building and multi-floor indoor localization
  fingerprint database widely used in the literature---is known to have the
  issue of uneven spatial distribution of RPs over buildings and floors. Data
  augmentation has been proposed as a feasible solution to not only improve the
  smaller number and the uneven distribution of RPs in the existing fingerprint
  databases but also reduce the labor and time costs of constructing new
  fingerprint databases. In this paper, we propose the multidimensional
  augmentation of fingerprint data for indoor localization in a large-scale
  building complex based on Multi-Output Gaussian Process (MOGP) and
  systematically investigate the impact of augmentation ratio as well as MOGP
  kernel functions and models with their hyperparameters on the performance of
  indoor localization using the UJIIndoorLoc database and the state-of-the-art
  neural network indoor localization model based on a hierarchical Recursive
  Neural Network (RNN). The investigation based on experimental results suggests
  that we can generate synthetic RSSI fingerprint data up to ten times the
  original data---i.e., the augmentation ratio of 10---through the proposed
  multidimensional MOGP-based data augmentation without significantly affecting
  the indoor localization performance compared to that of the original data
  alone, which extends the spatial coverage of the combined RPs and thereby
  could improve the localization performance at the locations that are not part
  of the test dataset.
\end{abstract}


\IEEEpeerreviewmaketitle

\section{Introduction}
\label{sec:introduction}
%%%
\IEEEPARstart{A}{s} the demand for location-based service (LBS) ever increases,
localization based on various wireless technologies is under extensive research
and development. Global Navigation Satellite System (GNSS) provides reliable,
real-time kinematic positioning and navigation in an outdoor environment, where
it takes only a few seconds to initialize and provide up to centimeter-level
accuracy~\cite{leandro2011rtx}.

In an indoor environment, however, the response time and accuracy of GNSS are
inadequate due to the blockage, attenuation, and scattering of satellite
signals by the obstacles inside and outside buildings~\cite{GPS}. At present,
indoor localization technologies are mainly based on
infrared~\cite{kemper2008challenges}, ultrasonic~\cite{kim2008advanced}, Ultra
Wide Band (UWB)~\cite{poulose2020uwb}, ZigBee~\cite{sugano2006indoor},
Bluetooth~\cite{altini2010bluetooth} and Wi-Fi~\cite{DNN}.

Note that, as modern buildings are already equipped with a large amount of
Wi-Fi infrastructure, indoor localization based on Wi-Fi technology does not
incur additional infrastructure overhead. Wi-Fi-based indoor localization
methods can be grouped into two, i.e., those based on \textit{ranging} and
\textit{location fingerprinting}. The ranging-based methods calculate the
distance between a user and Access Points (APs) based on received signal
measurements---e.g., angles in Angle of Arrival (AOA), and arrival times and
their differences in Time of Arrival (TOA) and Time Difference of Arrival
(TDoA)~\cite{yassin2016recent,survey}---to estimate a user's location via
multilateration, which requires the exact locations of APs in advance and, if
time measurements are involved, puts strict requirements on time
synchronization among all devices. The fingerprinting-based methods, on the
other hand, estimate a user's location by comparing the location fingerprint
like Received Signal Strength (RSS) or Received Signal Strength Indicator
(RSSI) measured at the user's current, unknown location during the online phase
with those pre-collected during the offline phase at known Reference Points
(RPs) in a location fingerprint database based on localization algorithms such
as Recursive Neural Network (RNN)~\cite{2021hierarchical} and
$k$-Nearest-Neighbor (kNN)~\cite{knn}, which, unlike the ranging-based methods,
does not require the locations of APs and strict time synchronization among the
devices. Their localization performance, however, could be significantly
affected by
% Therefore, there are some ambiguous points in the location
% determination, and its RSSI is similar to other RPs, which causes high
% estimation errors. This places a high demand on
the number and the coverage of the location fingerprints measured at the RPs in
the database, especially for a large-scale building complex~\cite{MOGP}.
% There are two phases in the indoor localization based on Wi-Fi fingerprinting:
% During the first \textit{offline phase}, an RSSI fingerprint database is
% constructed based on their measurements at RPs. The second phase is online
% localization matching, which compares the real-time localization information
% with the fingerprints in the database to determine the current spatial
% localization by localization algorithms such as Recursive Neural Network
% (RNN)~\cite{2021hierarchical} and $k$-Nearest-Neighbor (kNN)~\cite{knn}. The
% fingerprint database collected during the offline phase will directly affect the
% accuracy of the indoor localization based on Wi-Fi fingerprinting.

In fact, the uneven spatial distribution of RPs is a major issue among the
publicly-available location fingerprint databases like UJIIndoorLoc~\cite{UJI},
TUT~\cite{tut}, and WicLoc~\cite{database3}; in the case of the UJIIndoorLoc,
which is the mostly widely used multi-building and multi-floor RSSI database
and becomes a benchmark in the literature, the numbers of RPs are significantly
different for spaces with similar area, and many fingerprint samples have
spatial coordinates nearly identical to one another, indicating repeated
samplings at the same RPs. These problems result in an inadequate spatial
representation of data points and incomplete radio maps, which will be
discussed in detail in Section~\ref{sec:experimental_result}.
% The second issue is the lack of diversity in the database which makes it
% difficult to replicate. Most databases contain only a limited number of rooms,
% with small measurement areas and simple building structures. And public
% databases that take into account multi-building, multi-floor scenarios are
% more limited. Most researchers tend to create a small measurement area
% containing a limited number of APs, so that an ideal measurement environment
% is different from the actual localization where the localization service is
% provided.  Also, due to the lack of internal structural information of the
% buildings like floor plans, the researcher could only default to the database
% of different floors of the same building having the same layout.

To address these issues in fingerprint databases for large-scale multi-building
and multi-floor indoor localization, we propose the multidimensional
augmentation of fingerprint data based on Multi-Output Gaussian Process (MOGP)
in this paper. The proposed multidimensional fingerprint data augmentation can
improve the spatial coverage of data points of existing databases by generating
synthetic fingerprint data at additional RPs without loss of localization
accuracy. It could also reduce the labor and time costs of constructing new
databases using well-prepared but much reduced number of RPs, which could also
address the issue of difficult measurements related with complex building
structures.

The rest of the paper is organized as follows: In
Section~\ref{sec:related_work}, we first review the dominant methods in data
augmentation in general and proceed to the review of methods specific to indoor
localization. In Section~\ref{sec:method}, we propose the fingerprint data
augmentation for large-scale multi-building and multi-floor indoor localization
based on MOGP and discuss the details of the proposed algorithm including the
selection of the kernel function. Section~\ref{sec:experimental_result}
presents the results of our investigation of the impact of MOGP kernel
functions and models with their hyperparameters and augmentation ratio on the
performance of indoor localization using the UJIIndoorLoc database and the
state-of-the-art neural network indoor localization model based on a
hierarchical RNN~\cite{2021hierarchical}. Section~\ref{sec:conclusions}
concludes our work in this paper.

\section{Related Work}
\label{sec:related_work}
%%%
In this section we briefly review the basic ideas of data augmentation in
different research areas and the implementation of data augmentation specific to
indoor localization.

\subsection{Data Augmentation}
\label{sec:data_aug}
%%%
The success of Machine Learning (ML) algorithms highly depends on the existence
of a large amount of datasets, but the collection of datasets, especially
labeled ones for supervised learning, could be a challenging task in
applications such as large-scale invasive examinations in medical
testing~\cite{medical_image_GAN,hussain2017differential} and multi-building and
multi-floor indoor localization for a large-scale building complex~\cite{FASR}
due to the issues of privacy and the high labor and time costs. Data
augmentation has become a viable solution in this regard and applied widely to
the categorization of images~\cite{image_aug_survey} and
texts~\cite{wei2019eda}.

Image-based data augmentation algorithms can be grouped into two, i.e.,
image-processing-based and ML-based data augmentation: Image-processing-based
data augmentation utilizes image processing techniques such as geometric
transformations, flips, color transformations, cropping, noise and injection to
augment the data~\cite{image_aug_survey}. In the case of ML-based data
augmentation, advanced ML algorithms like deep neural networks are used; a
notable example is Generative Adversarial Networks (GANs), which emerge as a
representative approach to data augmentation using deep learning and have found
a wide range of applications in areas such as medical
imaging~\cite{medical_image_GAN} and urban traffic
control~\cite{transportation_gan}.

\subsection{Indoor Localization Data Augmentation}
\label{sec:loc_data_aug}
%%%
RSSI or RSS values can be converted into a grayscale map or plotted as a radio
map, enabling the application of the image-processing-based or ML-based data
augmentation techniques mentioned in Section~\ref{sec:data_aug}.

Rashmi Sharan Sinha et al. converted a file containing 256 RSSI values
into a $16{\times}16$ image as input to a Convolutional Neural Network
(CNN)~\cite{electronics9050851,sinha2019data}. Tian Lan et al. proposed a
super-resolution-based fingerprint augmentation framework to achieve
interconversion between fingerprint data and fingerprint images~\cite{FASR}. 

Direct augmentation of indoor localization data using ML algorithms such as GAN
are popular nowadays. Wafa Njima et al. used a Selective-GAN to augment the
UJIIndoorLoc database, and the localization prediction in the offline phase is
demonstrated to significantly improve the localization
accuracy~\cite{njimaIndoorLocalizationUsing2021}. Hilal et al.  proposed
DataLoc+, a room-level data augmentation technique inspired by the dropout
technique to prevent overfitting~\cite{DataLoc+}. Rizk et al. used deep learning
to implement data augmentation in cellular-based
localization~\cite{Cellular-based}. In~\cite{aug_GPR} and~\cite{Kriging}, the
researchers used Single-Output Gaussian Process (SOGP) regression, also called
\textit{Kriging} in geostatistics, to augment the indoor localization data with
single building and single floor.

Note that there was no prior work on the use of MOGP to explicitly exploit the
correlation among observations from multiple APs in multi-building and
multi-floor indoor localization and investigate an optimal way of RSSI data
augmentation based on MOGP, which is the major contribution of our work in this
paper.

\section{Multidimensional Fingerprint Data Augmentation Based on MOGP}
\label{sec:method}
\begin{figure}[htb]
  \centering%
  \includegraphics[width=\linewidth, trim=0 0 0 0,clip=true]{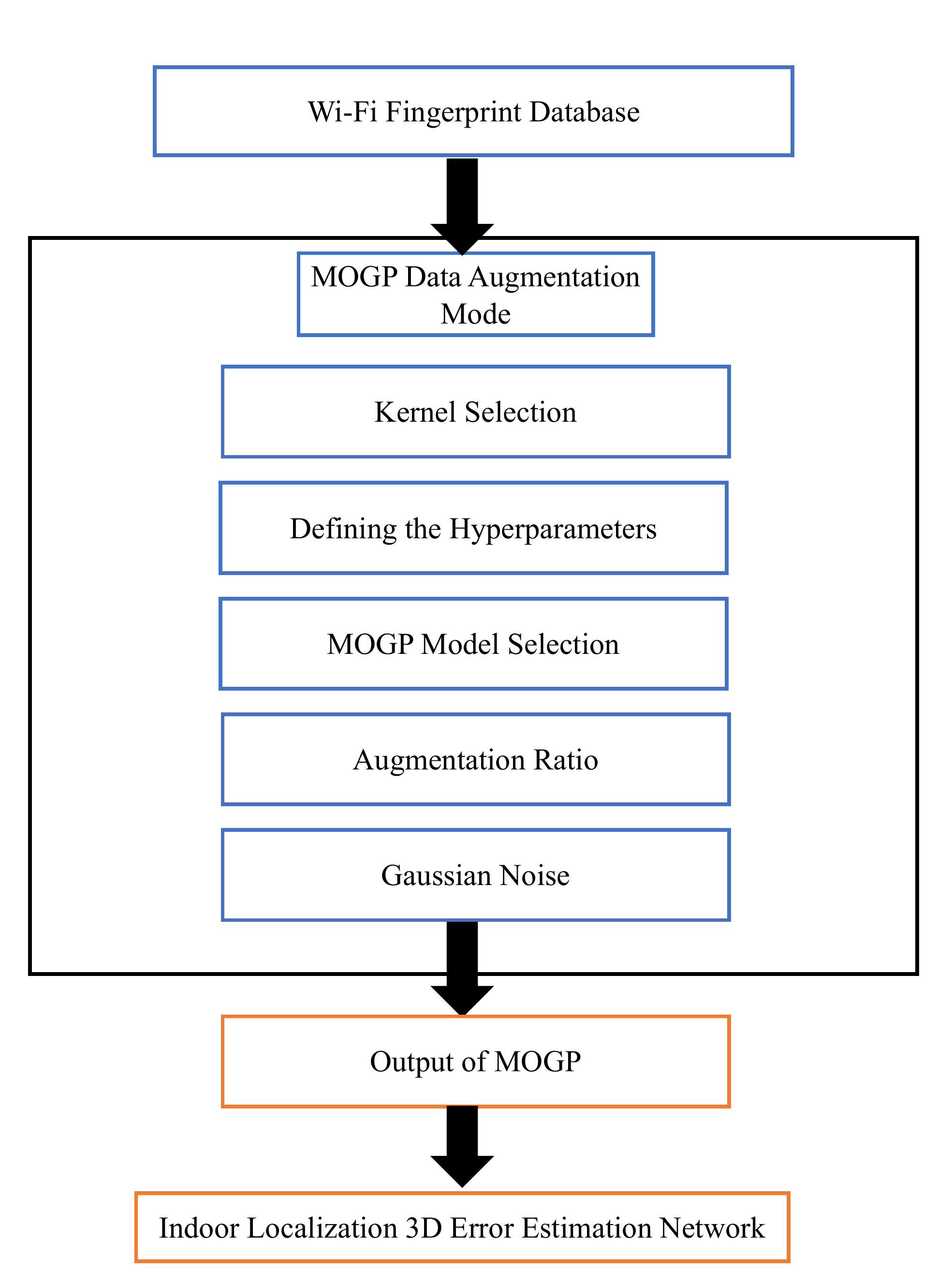}
  \caption{An overview of multidimensional fingerprint data augmentation based
    on MOGP.}
  \label{fig:system_overview}
\end{figure}

%%%
Fig.~\ref{fig:system_overview} provides an overview of the proposed
multidimensional fingerprint data augmentation based on MOGP. MOGP-based data
augmentation algorithm, in the black box, has five steps and it belongs to the
offline data collection and processing phase, and its output will be the input
of the online localization estimation and prediction network, which is the RNN
used in this study. It is first necessary to decide on the kernel function with
its hyperparameters, then the system provides two different MOGP models, the
augmentation ratio directly determines the amount of augmented data, and after
adding Gaussian noise the output of MOGP will be obtained.
%%%
%%%

\subsection{From Single-Output to Multi-Output Gaussian Process}
\label{sec:mogp}
%%%
For multi-building, multi-floor indoor localization scenarios, we can define
RSSI matrix for $N$ Wireless Access Points (WAPs) and $M$ RPs as follows:
%%%
\begin{equation}
  \mathbf{RSSI}=[\mathbf{RSSI_{1}},\mathbf{RSSI_{2}},\cdots,\mathbf{RSSI_{N}}],
  \label{equ:rssi_matrix}
\end{equation}
where $\mathbf{RSSI_{n}}$ is a column vector with RSSI measurement of WAP$n$.
%%%
For the UJIIndoorLoc database, a total of 520 different WAPs measurements
yielded 19938 RPs~\cite{UJI}, and \eqref{equ:rssi_matrix} includes the RSSI
values of each RP. Another input
matrix~\eqref{equ:geographic_information_matrix} includes the geographic
information of RPs, $log$ and $lat$ for longitude and latitude, respectively.
$b$ denotes building number, $f$ denotes floor. The input of the model should
be the concatenation of the matrices $\bf{RSSI}$ and $\bf{RP}_{GI}$, the latter
of which is for the geographic information of RPs and defined as
%%%
\begin{equation}
  \bf{RP_{GI}} = [RP_{i,log}, RP_{i,lat}, RP_{i,b} , RP_{i,f}],
  % \begin{bmatrix}
  % RP_{i,x} & RP_{i,y} & RP_{i,b} & RP_{i,f} \\
  % RP_{1,x} & RP_{1,y} & RP_{1,b} & RP_{1,f} \\
  % \vdots & \vdots & \vdots & \vdots \\
  % RP_{M,x} & RP_{M,y} & RP_{M,b} & RP_{M,f} \\
  % \end{bmatrix}%\in \mathbb{R}^{M}
  \label{equ:geographic_information_matrix}
\end{equation}
%%%
where $i{\in}[1,M]$.

SOGP can be defined as follows:
%%%
\begin{equation}
  \textit{f} (\mathbf{x})\sim \mathcal{GP}(m(\mathbf{x}),k(\mathbf{x},\mathbf{x})),
  \label{equ:sogp}
\end{equation}
%%%
where $m(\textbf{x})$ is a mean function normally set to $0$, and
$k(\textbf{x},\textbf{x})$ is a kernel function also called a covariance
function and $\Omega_{f}$ defines the output space of SOGP. SOGP can be interpreted as the union of a series of random variables
about a continuous domain, and for each spatial point the random variables obey
a Gaussian distribution, and since the main purpose of the study is indoor
localization the situation in space is discussed, ignoring the time dimension.
Hence taking the UJIIndoorLoc database as an example, each RP corresponds to
520 identified SOGPs, as the database records 520 WAPs. For SOGP, the kernel
function of the model only stores the RSSI value and geographic information of
current WAP, hence the input matrix of the SOGP is
\begin{equation}
  \mathbf{x} = \mathbf{[RSSI_{n},RP_{GI}]},
\end{equation}
where $n{\in}[1,{\cdots},520]$.  As for output, given the relationships between
associated observations and the output,
%%%
\begin{equation}
  y_{t}=f_{t}(\mathbf{x})+\epsilon _{t},t\in1\leqslant t\leqslant T
  \label{equ:regression}
\end{equation}
%%%

\begin{figure}[htb]
  \setlength{\myvspace}{2mm}
  \begin{center}
    \includegraphics[angle=-90,width=.5\linewidth]{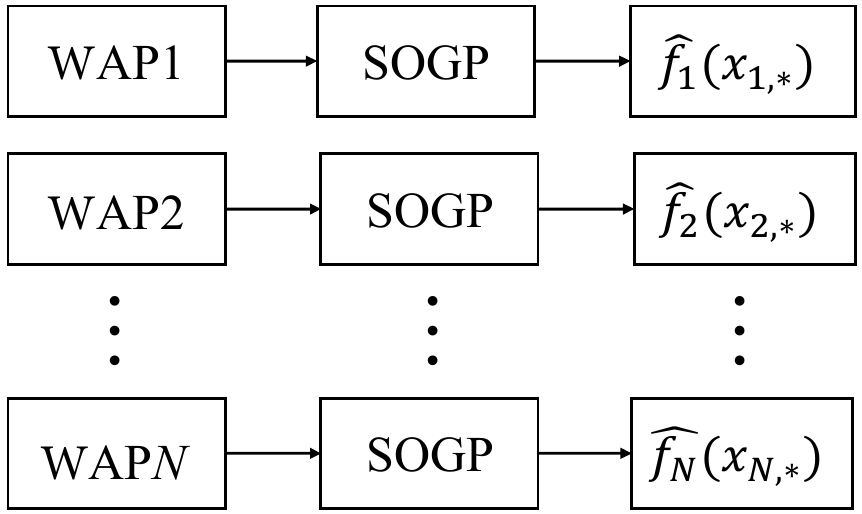}\\
    \vspace{\myvspace}
    {\scriptsize (a)}\\
    \includegraphics[angle=-90,width=.6\linewidth]{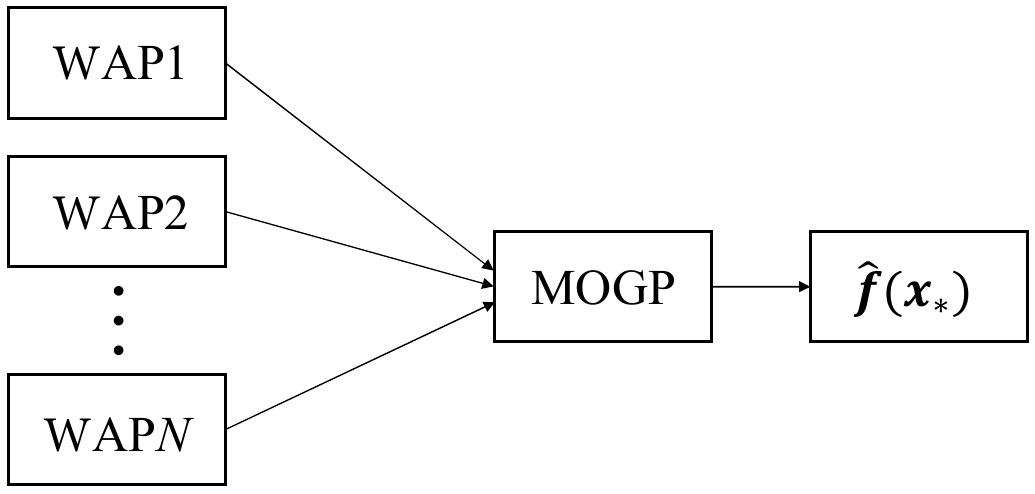}\\
    \vspace{\myvspace} {\scriptsize (b)}
  \end{center}
  \caption{Block diagram of (a) SOGP and (b) MOGP from~\cite{MOGP}.}
  \label{fig:block_diagram}
\end{figure}

For MOGP model, it can be defined as,
\begin{equation}
  \text{MOGP}: \Omega_{d} \rightarrow  \{\Omega_{f_{1}}  \cdots \Omega_{f_{T}}\},
\end{equation}
where $ \Omega_{d}$ related to the d-dimensional input space, in this research is $[RSSI_{n}, RP_{GI}]$, and $\Omega_{f}$ is the output space of the $f_{t}(\textbf{x})$~\cite{remarks_GP}. The $T$ outputs obtained from MOGP can be used to classify the MOGP model according to whether the importance is equal or not, and this part will be discussed further in Section~\ref{sec:mogp_model}.
For indoor localization related studies, RSSI values measured by different WAPs
at the same RP are correlated, so a multi-input multi-output Gaussian Process
Regression model can be used to consider the correlation between different
WAPs. It is worth noting that both the
SOGP and MOGP models ignore their multi-input characteristics. The transition
from SOGP to MOGP for indoor localization studies requires making the
assumption that multi-outputs are correlated and of equal importance to each
other. In this context, it is possible to map data from multi-sequences of
multitasks into the same real-valued function space and to line them up, the
multi-output scenario is made to satisfy the definition of a Gaussian Process
which is the Gaussian measure is in the space where the real-valued stochastic
process is located and this is also considered as symmetric
MOGP~\cite{remarks_GP}. The multi-task here can be interpreted as including
regression of RSSI based on latitude and longitude, regression of RSSI based on
floor coordinates, regression of RSSI based on building coordinates, and a
joint task of the above three tasks. Please note that the floor and building
coordinates here are not continuous values and cannot be used directly for MOGP
regression, and the detailed experimental setup will be explained in
Section~\ref{sec:experimental_result}. Fig.~\ref{fig:block_diagram} image
showing the difference between SOGP and MOGP in output and training.
%%%
%%%

\subsection{Different MOGP Models}
\label{sec:mogp_model}
%%%
The MOGP model can be simply divided into symmetric MOGP and asymmetric MOGP by
determining the importance of maintaining the same weights among
outputs~\cite{MOGPremarks}. Symmetric MOGP is characterized by the use of a
symmetric structure to preserve correlations between outputs and to ensure that
the outputs satisfy condition $\{f_{t}\}_{1\leqslant t\leqslant T}$ follows the
regression observations with independent and identically distributed Gaussian
measurement noise, shown in (~\ref{equ:regression}).

The correlation information is stored in an integrated kernel function and all
outputs share the same training information. The kernel function of the MOGP
can be defined as
%%%
\begin{equation}
  K_{M}(\mathbf{x},\mathbf{x}')=\begin{bmatrix}
    k_{11}(\mathbf{x},\mathbf{x}^{'}) & \cdots & k_{1T}(\mathbf{x},\mathbf{x}^{'}) \\
    k_{21}(\mathbf{x},\mathbf{x}^{'}) & \cdots & k_{2T}(\mathbf{x},\mathbf{x}^{'}) \\
    \vdots                            & \ddots & \vdots                            \\
    k_{T1}(\mathbf{x},\mathbf{x}^{'}) & \cdots & k_{TT}(\mathbf{x},\mathbf{x}^{'}) \\
  \end{bmatrix}.
  \label{equ:mogp_kernel_matrix}
\end{equation}
%%%
The latent function $u(\textbf{x})$ in \eqref{equ:ux} is assumed to be a
Gaussian Process satisfying zero mean function and kernel function as
$k(\textbf{x},\textbf{x}^{'})$.
%%%
\begin{equation}
  \textit{u} (\textbf{x})\sim \mathcal{GP}(0,k_{q}(\textbf{x},\textbf{x}^{'})).
  \label{equ:ux}
\end{equation}
%%%
Thus for a determined kernel function can sample $R$ latent functions from its
corresponding MOGP model. The $R$ latent functions obtained by simultaneous
sampling all obey a GP, which is
%%%
\begin{equation}
  u^{1}(\mathbf{x}),u^{2}(\mathbf{x}),{\cdots},{u}^{R}(\mathbf{x}){\sim}\mathcal{GP}(0,k(\mathbf{x},\mathbf{x}^{'}))
  \label{equ:sampled_ux}.
\end{equation}
%%%
For this series of latent functions $u(x)$, the purpose of the line
transformation is to obtain $T$ outputs, $f_{1}(x),f_{2}(x),{\cdots},f_{T}(x)$,
and the generic format of the output function $f(x)$ can be written as,
%%%
\begin{equation}
  f_{t}(\mathbf{x})=a^{1}_{t}u^{1}(\mathbf{x})+a^{2}_{t}u^{2}(\mathbf{x})+\cdots+a^{R}_{t}u^{R}(\mathbf{x}),t\in[1,T].
  \label{equ:outputs_icm}
\end{equation}
%%%
The above form of sampling the output using one given kernel function is known
as Intrinsic Coregionalization Model (ICM). If the number of kernel functions
$Q$, large than one, so that the model is the Linear Model of Coregionalization
(LMC). The LMC samples a total of $Q$ kernel functions
$k_{q}(x,x^{'})\,q{\in}[1,Q]$, each yielding $R$ latent functions, and obtains
$T$ output functions by means of a linear combination. The number of kernel
functions $Q$ influences the expressiveness of the model to some extent.
Therefore, \eqref{equ:outputs_icm} can be rewritten as follows:
%%%
\begin{equation}
  f_{t}(\mathbf{x})=\sum_{q=1}^{Q}\sum_{r=1}^{R}a_{t,q}^{r}u_{q}^{r}(\mathbf{x}).
  \label{equ:outputs_lmc_1}
\end{equation}
%%%
Some researchers have suggested using $Q{=}2$~\cite{nguyen2014collaborative} or
$Q{=}T$~\cite{fricker2013multivariate} to improve the flexibility of the model
and its ability to describe differences in the data.

\subsection{Kernel Function}
\label{sec:kernel}
Given that a GP model can be uniquely determined by a mean function, which is
usually set to zero, and a kernel function, this section will focus on the
characteristics of different kernel functions. The field of ML does not make a
very clear distinction between kernel functions and covariance functions; both
can be regarded as generalized descriptions of distances. Also, as kernel
functions are linear in character, new kernel functions can be constructed by
simple linear combinations, as
%%%
\begin{equation}
  kernel_{new} = \sum_{n=1}^{\infty }A_{n}kernel_{n}.
  \label{equ:outputs_lmc}
\end{equation}
%%%
Thus there is challenging to exhaust all cases by enumeration for linear
combinations of arbitrary kernel functions, and some researchers have attempted
to discuss the performance of some simple combinations in indoor localization
data augmentation, such as the use of compound kernel functions, linear
combination of noise, constant and linear kernel~\cite{aug_GPR}. Note that this
is a SOGP regression forecast based on a single WAP. Therefore in this study we
focus on the effectiveness of MOGP using a single kernel function in the
augmentation of indoor localization databases in large scale multi-floor
building complexes. The corresponding kernel function for MOGP can be
constructed by extending the kernel function for SOGP so that it adds an
additional discrete input dimension~\cite{duvenaud-thesis}.

The most common kernel function is the Radial Basis Function (RBF) or what is
known as the Gaussian kernel function, which is defined by
%%%
\begin{equation}
  k_{RBF}(\mathbf{x},\mathbf{x}')=\sigma^{2}exp(-\frac{(\mathbf{x}-\mathbf{x}')^{2}}{2l^{2}}),
  \label{equ:rbf}
\end{equation}
%%%
where $\sigma^{2}$ is the variance describing the average distance from the mean
and $l$ is the length-scale representing the spread of the covariance. In most
cases the RBF kernel function fits well and the correlation between individual
data points in the domain is generally considered to decay smoothly with
increasing distance~\cite{RBF}. However, in some cases such a pre-determined
data correlation satisfying such a smooth decay is not true; in the case of a
unit step like signal, for example, the RBF kernel function does not capture the
characteristics of the signal at the moment of the jump accurately and tends
more to amplify the time of change of the signal.
%%%
\begin{equation}
  k_{RQ}(\mathbf{x},\mathbf{x}')=\sigma ^{2}exp(1+\frac{(\mathbf{x}-\mathbf{x}')^{2}}{2\alpha l_{RQ}^{2}})^{-\alpha },
  \label{equ:rq}
\end{equation}
%%%
where Rational Quadratic (RQ) kernel function is the mixture of the RBF kernel
with different length-scale $l$~\cite{wilson2013gaussian}. When
$\alpha{\to}\infty $, the RQ becomes the RBF kernel
function~\cite{duvenaud2014automatic}. However, it does not solve the problem
of excessive smoothness very well.

The use of the Matern class of kernel functions goes some way to alleviating
the problem of over-smoothing at signal jump moments~\cite{Matern}. The Matern
family of kernel functions can be defined by
%%%
\begin{equation}
  k_{Matern}^{v}(\mathbf{x},\mathbf{x}')= \frac{2^{1-v}}{\Gamma (v)}(\frac{\sqrt{2v\left| \mathbf{x}-\mathbf{x}'\right|}}{l})^{v}K_{v}(\sqrt{2v\left| \mathbf{x}-\mathbf{x}'\right|}),
  \label{equ:matern}
\end{equation}
%%%
where $K_{v}$ is modified Bessel function, by changing the parameter
$v{=}d{+}\frac{1}{2}$, where $d$ is the order of a polynomial function, the
problem of over-smoothing of the RBF kernel function in the signal mutations
region can be mitigation. While $v{\to}\infty $, the Matern kernel function has
the same structure as RBF kernel function. Hence, the general case $v$ takes
the values $\frac{3}{2}$ or $\frac{5}{2}$, and Matern32 will be coarser for
Matern52. Another solution to the over-smoothing of the RBF kernel function is
to replace the quadratic Euclidean distance with the absolute distance, which
is the Ornstein-Uhlenbeck (OU) kernel function defined in \eqref{equ:ou}. The
OU kernel function is also essentially a Matern kernel function when $v$ takes
the value $1$~\cite{williams2006gaussian}.
%%%
\begin{equation}
  k_{OU}(\mathbf{x},\mathbf{x}')=exp(-\frac{\left\| \mathbf{x}-\mathbf{x}'\right\|}{l})
  \label{equ:ou}
\end{equation}
%%%
The differences between several kernel functions are compared above mainly in
terms of smoothness, with other aspects being ignored, and their impact and
performance on the enhancement of indoor positioning data is compared in detail
in Section~\ref{sec:experimental_result}.

\section{Experimental Results}
\label{sec:experimental_result}
%%%
The publicly-available UJIIndoorLoc database is used for the experiments, which
has one five-floor and two four-floor buildings~\cite{UJI}. As for the
evaluation of the localization performance of the proposed multidimensional
fingerprint data augmentation based on MOGP, which is presented in
Section~\ref{sec:online-evaluation}, we use the state-of-the-art neural network
model based on the hierarchical RNN specifically designed for large-scale
multi-building and multi-floor indoor localization ~\cite{2021hierarchical}. We
also provide the results of our investigation of the impact of MOGP kernel
functions on RSSI radio maps in Section~\ref{sec:offline-evaluation}.

\subsection{Online Evaluation}
\label{sec:online-evaluation}
%%%
During the online evaluation, we investigate the impact of different MOGP kernel
functions and models with their hyperparameters on the indoor localization
performance of the proposed data augmentation. We also explore the ratio of data
augmentation to have an insight in generating synthetic RSSI data without
causing overfitting.
% We use the RNN to estimate the position and calculate the error to judge the
% performance of the data augmentation.

\begin{figure}[htb]
  \centering%
  \includegraphics[angle=-90,width=.9\linewidth]{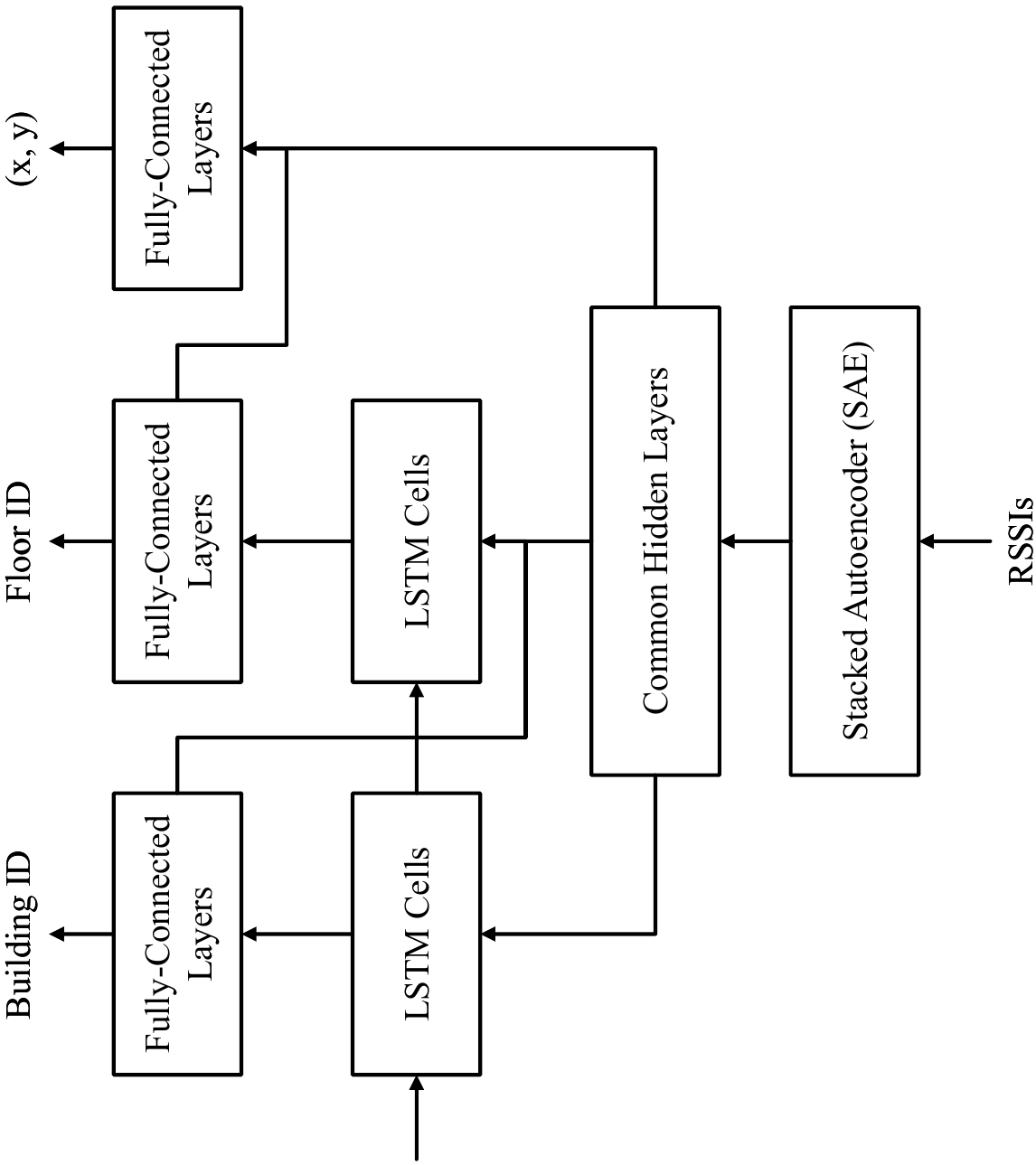}
  \caption{Network architecture of the RNN indoor localization model with LSTM
    cells~\cite{2021hierarchical}}.
  \label{fig:rnn}
\end{figure}

We select the RNN model in~\cite{2021hierarchical} with Long Short-Term Memory
(LSTM) cells to investigate the effects of the MOGP-based augmentation on the
localization performance. The MOGP regression is implemented based on
GPy---i.e., a GP framework in Python developed at the Sheffield machine learning
group~\cite{GPy}---following the steps outlined in Section~\ref{sec:mogp}. As
shown in Fig.~\ref{fig:rnn}, the Stacked AutoEncoder (SAE) of the RNN model
consists of three hidden layers of 256, 128, and 64 nodes, which is followed by
two common hidden layers with 128 and 128 nodes.
%%%
%%%

For building and floor classifiers, we have two stacked LSTM cells followed by
two Fully-Connected (FC) layers of 32 nodes and 1 output node. The position
estimator consists of three FC layers of 512 and 512 nodes and 2 output nodes
for two-dimensional localization coordinates~\cite{2021hierarchical}. We apply
\textit{early stopping} with a patience of 20 for the coordinate estimation
model and 40 with \textit{save best only} functions activated for the
building/floor classification model. Table~\ref{tbl:hyperset} summarizes the
key RNN parameter values for the experiments.
%%%
\begin{table}[htb]
  \caption{RNN parameter values.}
  \label{tbl:hyperset}
  \centering%
  \begin{tabular}{ll}
    \hline
    Parameter                               & Value      \\
    \hline
    SAE Hidden Layers                       & 256-128-64 \\
    SAE Activation                          & ReLu       \\
    SAE Optimizer                           & Adam       \\
    SAE Loss                                & MSE        \\
    Common Hidden Layers                    & 128-128    \\
    Common Activation                       & ReLu       \\
    Common Dropout                          & 0.2        \\
    Common Loss                             & MSE        \\
    LSTM Cells                              & 256-512    \\
    LSTM Activation                         & ReLu       \\
    LSTM Optimizer                          & Adam       \\
    LSTM Loss                               & MSE        \\
    Building/Floor Classifier Hidden Layers & 32-1       \\
    Building/Floor Classifier Activation    & MSE        \\
    Building/Floor Classifier Optimizer     & Adam       \\
    Building/Floor Classifier Dropout       & 0.2        \\
    Building/Floor Classifier Loss          & ReLu       \\
    Position Estimator Hidden Layers        & 512-512-2  \\
    Position Estimator Activation           & MSE        \\
    Position Estimator Optimizer            & Adam       \\
    Position Estimator Dropout              & 0.1        \\
    Position Estimator Loss                 & tanh       \\
    \hline
  \end{tabular}
\end{table}
%%%

As performance metrics for the localization performance, we use \textit{building
  hit rate} and \textit{floor hit rate} defined as a rate of correct
identification of building Identifier (ID) and that of floor ID, respectively,
and \textit{3D error} defined as the mean of three-dimensional Euclidean
distances between estimated and correct locations~\cite{EvAAL}.

\begin{figure}[htb]
  \centering%
  \includegraphics[width=.8\linewidth,trim=160 40 160 85,clip]{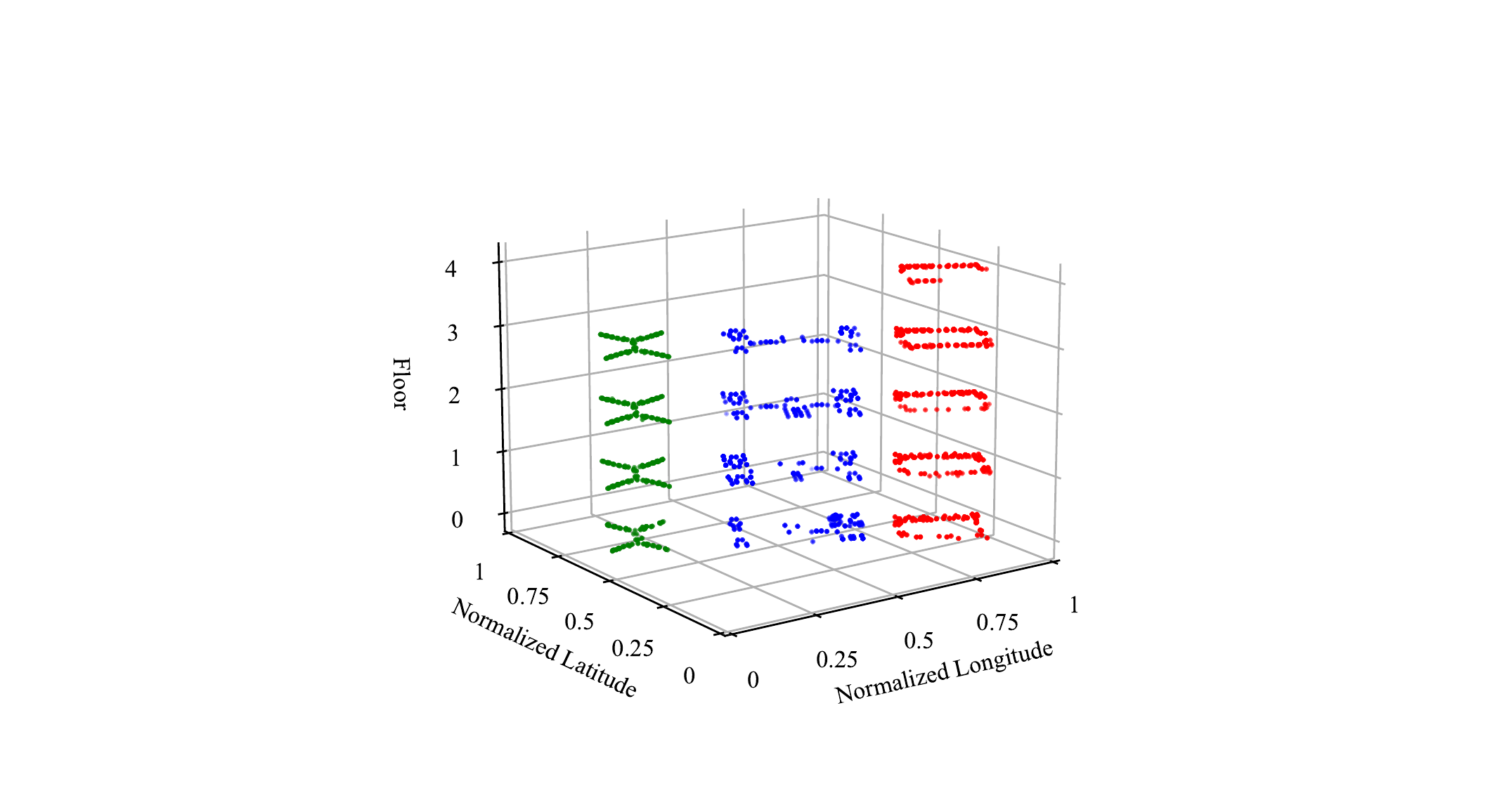}
  \caption{Spatial distribution of database UJIIndoorLoc. }
  \label{fig:ujidata}
  %  %%% 
  % \comment{Leo}{The labels of the x and y coordinates are not clear;
  % We need to clarify the units of the x and y coordinates are normalized:
  % i.e., 'When visualizing data points, we use normalized coordinates to
  % describe them more clearly since the difference between them is small
  % compared to the vast upper and lower bounds of the original data's longitude
  % and latitude (x and y coordinates).'}
  % \response{Tim}{Note that the axes in Fig. 5, 6, and 7 are not normalized, so
  % if the above explanation is added here we may need to redraw Figures 5, 6,
  % and 7 to ensure that they are all in normalized coordinates, so my
  % suggestion is to set him aside or add the following text.  "Fig. 4 uses
  % normalized coordinates for a wide area of the three buildings in order to
  % represent the square. The later figures still use real coordinate because
  % the area they cover is limited."}
  % %%% 
\end{figure}

\begin{figure}[htb]
  \centering%
  \includegraphics[width=\linewidth,trim=205 0 220 0,clip=true]{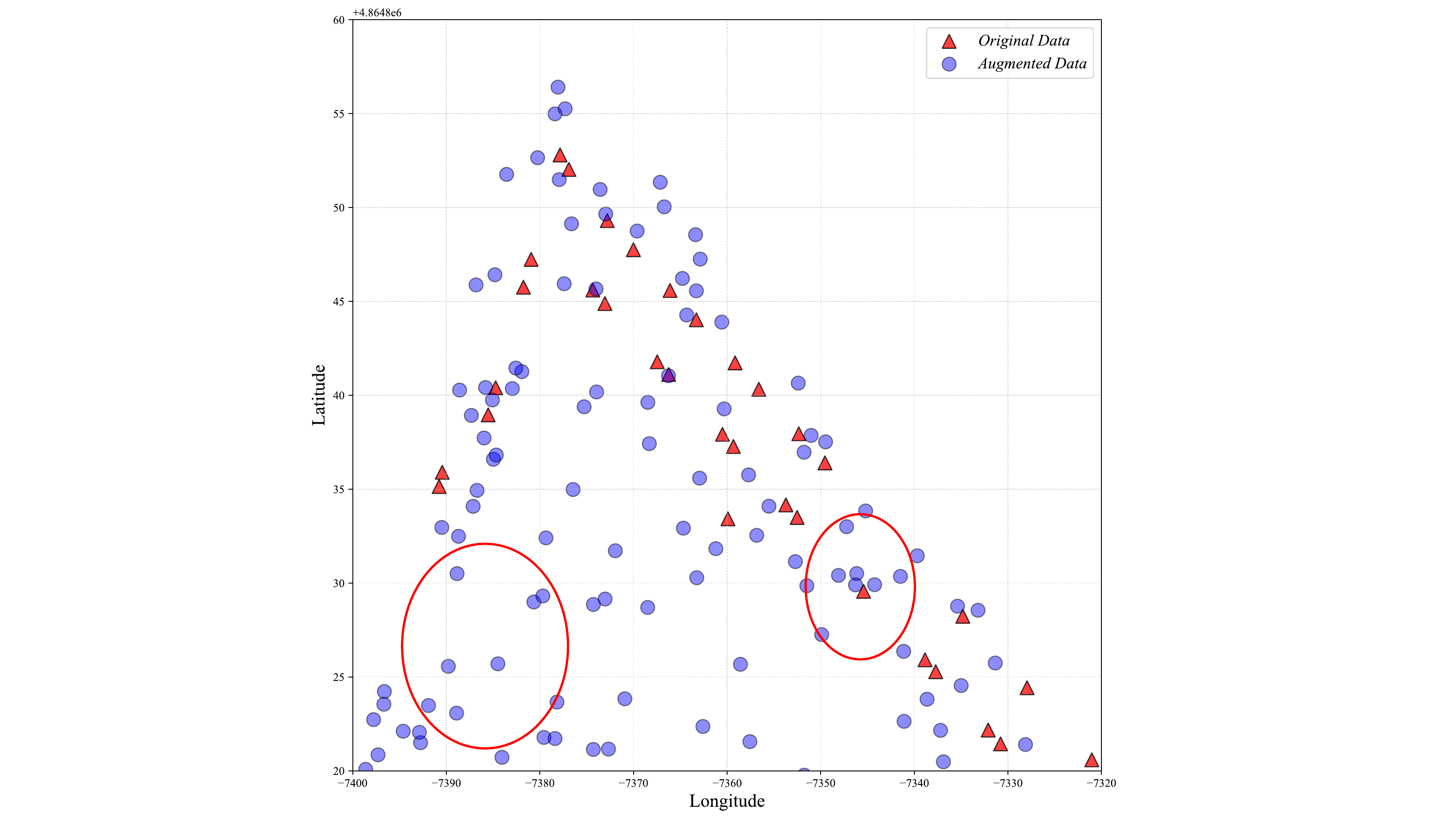}
  \caption{Spatial distribution and coverage of the original and the augmented
    data for the corner of the 4th floor of building~2 of the UJIIndoorLoc
    dataset.}
  \label{fig:data}
\end{figure}

Fig.~\ref{fig:ujidata} illustrates the spatial distribution of RPs over three
buildings in normalized coordinates, which clearly shows the poor spatial
coverage of RPs for building 1 and 2 indicated by blue and red dots,
respectively.
%%%
%%%
Table~\ref{tbl:rp-number} summarizes the
statistics of the number of RPs on different floors of different
buildings.
%%%
\begin{table}[htb]
  \centering
  \caption{Statistics on the number of RPs on different floors of different
    buildings of UJIIndoorLoc database.}
  \label{tbl:rp-number}
  \begin{tabular}{cccc}
    \hline
            & Building 0 & Building 1 & Building 2 \\ \hline
    Floor 0 & 1059       & 1368       & 1942       \\
    Floor 1 & 1356       & 1484       & 2162       \\
    Floor 2 & 1443       & 1396       & 1577       \\
    Floor 3 & 1391       & 948        & 2709       \\
    Floor 4 & None       & None       & 1102       \\
    Total   & 5249       & 5196       & 9492       \\ \hline
  \end{tabular}
\end{table}
%%%
Within the same building with similar building structure and spatial area, the
difference in the numbers of RPs on different floors is obvious, especially for
building 2 as shown in Fig.~\ref{fig:ujidata}. Fig.~\ref{fig:data} illustrates
one possible way of sampling the latitude and longitude using a Gaussian
distribution to determine the physical spatial coordinates of the augmented data
taking the north-west corner of floor 4 of building 2 as an example.
%%%
%%%
The triangles and the circles indicate the RPs in the original database and the
augmented RPs, respectively. From the figure, we can clearly see the cavities
caused by the absence of measurement data, which are highlighted by red circles;
according to the original database, these RPs are distributed within the
corridor, and the augmented data successfully fill the cavities in the original
database.

\subsubsection{Impact of Kernel Functions}
\label{sec:kernel-performance}
%%%
The impact of different kernel functions in data augmentation on the
localization performance is discussed in Section~\ref{sec:kernel}, and the
results of its experimental verification are summarized in
Table~\ref{tbl:different_kerenls}, where we use ICM and set the ratio of data
augmentation to 1 with the variance of 1 and the length-scale of 10. Given the
parameter settings, Matern52 provides optimal results for a single kernel
function. Visualizing the results of the data augmentation (i.e., numerical RSSI
values), we found that the MOGP model narrows the range of fluctuations of the
original data regardless of the kernel function chosen. This excludes specific
extreme value points, which could result in data dilution when mixed with the
original database.
%%%
\begin{table}[htb]
  \caption{Impact of kernel functions on indoor localization.}
  \label{tbl:different_kerenls}
  \centering
  \begin{tabular}{cccccc}
    \hline
    Kernel Function  & RBF  & RQ   & Matern32 & Matern52 & OU   \\ \hline
    3D Error {[}m{]} & 9.04 & 9.26 & 8.88     & 8.70     & 8.93 \\ \hline
  \end{tabular}
\end{table}
%%%

\subsubsection{Impact of the Hyperparameters of Kernel Functions}
\label{sec:kernel-hyperparameters}
%%%
A kernel function consists of two hyperparameters of a variance $\sigma^{2}$
(also called scale factor) and a length-scale $l$. Table~\ref{tbl:variance}
shows the impact of the variance on the localization performance. The variance
$\sigma^{2}$ scales the kernel function and controls the average distance from
the mean function. A variance greater than one magnifies the change in kernel
function to the mean, and this will amplify some of the small changes between
the data. The intuitive application to indoor localization data augmentation is
the ability to fit local extrema for a single WAP. To a certain extent, a larger
variance can solve the problem of too smooth a kernel function for data
augmentation, whose effect, however, is very limited; selecting appropriate
kernel functions for data and application scenarios remains an open issue.
%%%
\begin{table}[htb]
  \caption{Impact of variance on indoor localization.}
  \label{tbl:variance}
  \centering
  \begin{tabular}{cccc}
    \hline
    Variance      & 0.1  & 1    & 10   \\ \hline
    3D Error  [m] & 8.96 & 8.70 & 8.82 \\ \hline
  \end{tabular}
\end{table}
%%%

In general, the length-scale $l$ controls the extrapolation capability of the
model or defines the limiting distance to which the maximum predictable
belongs. If $l$ is set to low, the model will place extra emphasis on areas with
large data fluctuations. The large $l$ causes the model to lose its ability to
capture subtle changes, so it is recommended to use a relatively small $l$ in
indoor localization data augmentation. The result shown in
Table~\ref{tbl:length-scale} supports this conclusion.
%%%
\begin{table}[htb]
  \caption{Impact of length-scale on indoor localization.}
  \label{tbl:length-scale}
  \centering
  \begin{tabular}{cccc}
    \hline
    Length-scale & 1    & 10   & 100  \\ \hline
    3D Error [m] & 8.78 & 8.70 & 8.83 \\ \hline
  \end{tabular}
\end{table}
%%%

\subsubsection{Impact of MOGP Models: LMC v.s. ICM}
\label{sec:lmc-vs-icm}
%%%
The differences between LMC and ICM and the impact of the number of kernel
functions $Q$ on the performance of the model have been discussed in
Section~\ref{sec:mogp}. Table~\ref{tbl:model} shows the localization errors of
different MOGP models using Matern52 kernel function with both variance and
length-scale set to $1$ and augmentation ratio to 1. When $Q{\in}[2,4]$, there
was no statistically significant difference in the mean values after multiple
independent calculations, especially given the fluctuations in the localization
error estimates of the RNN. For the UJIIndoorLoc database only, the localization
error reaches a minimum when the number of kernel functions $Q$ is equal to the
number of outputs of the MOGP $T$. However, given the significant increase in
computation time as $Q$ increases for databases with a small number of APs,
$Q{=}2$ can be chosen to balance performance and efficiency when
augmenting. This finding is also in line with other researchers' suggestions for
the number of latent
processes~\cite{fricker2013multivariate,nguyen2014collaborative}.
%%%
\begin{table}[htb]
  \caption{Impact of MOGP model on indoor localization.}
  \label{tbl:model}
  \centering
  \begin{tabular}{ccc}
    \hline
    MOGP Model           & Numbers of Sample Q & 3D Error [m] \\ \hline
    ICM                  & 1                   & 8.70         \\ \hline
    \multirow{4}{*}{LMC} & 2                   & 8.60         \\
                         & 3                   & 8.58         \\
                         & 4                   & 8.61         \\
                         & T                   & 8.42         \\ \hline
  \end{tabular}
\end{table}
%%%

\subsubsection{Impact of Augmentation Ratio}
\label{sec:augmentation_ratio}
%%%
Given the uneven spatial distribution of measurement points in the original
database, and even the absence of measurements in some areas, it would be
valuable to discuss the scale of data augmentation. The augmentation ratio is
defined as
%%%
\begin{equation}
  r=\frac{\text{Number of Augmented Data}}{\text{Number of Original Data}}.
  \label{equ:ratio}
\end{equation}
%%%
Excessive data augmentation significantly increases the amount of augmented data
over the original data, which would cause overfitting the RNN and the ignorance
of the features of the original data. Also a small augmentation ratio may leave
the areas not covered by the original data still not covered by the augmented
data as well, making the RNN unable to capture the fingerprint features in that
areas during the localization estimation; it may also mark augmentation data as
noise. Table~\ref{tbl:r} shows the errors in the localization estimates for
different augmentation ratios. Unlike the previous experiments, the results in
Table~\ref{tbl:r} use LMC, and $T$ latent variables. The localization error can
therefore be taken to a minimum when the augmentation ratio $r{\approx}1$ is
applied. Also in the quest for maximum spatial coverage, the amount of data can
be expanded by a factor of 10 with a 95\% confidence interval, using the
unaugmented data as a benchmark.
%%%
\begin{table}[htb]
  \caption{Impact of augmentation ration on indoor localization.}
  \label{tbl:r}
  \centering
  \begin{tabular}{ccccccc}
    \hline
    Ratio        & 0.05 & 0.5  & 1    & 5    & 10   & Original~\cite{2021hierarchical} \\ \hline
    3D Error [m] & 9.44 & 8.93 & 8.59 & 8.68 & 8.74 & 8.62                             \\ \hline
  \end{tabular}
\end{table}
%%%

The optimal parameters of the MOGP-based multidimensional indoor localization
data augmentation algorithm for large-scale building complexes and its error
performance are given in a comprehensive discussion of the above parameters.
Table~\ref{tbl:result} summarizes the optimal results from~\cite{MOGP}.
%%%
\begin{table}[!htb]
  \caption{Multidimensional indoor localization errors and parameters~\cite{MOGP}.}
  \label{tbl:result}
  \centering
  \begin{tabular}{cc}
    \hline
    Parameter            & Value    \\ \hline
    Kernel Function      & Matern52 \\
    Variance             & 1        \\
    Length-scale         & 10       \\
    Model                & LMC      \\
    Number of Sample Q   & T        \\
    Augmentation Ratio   & 1        \\
    3D Error Mean [m]    & 8.59     \\
    3D Error Minimum [m] & 8.42     \\ \hline
  \end{tabular}
\end{table}
%%%

The Table~\ref{tbl:result_compare} shows a comparison of the results with other
participants in the EvAAL competition, but please note that this is not a fair
comparison as the UJIIndoorLoc test set is not accessible to non-participants.
Also optimal results are mentioned in earlier publications~\cite{MOGP}.
%%%
\begin{table}[!htb]
  \caption{Result Comparison~\cite{MOGP}}
  \label{tbl:result_compare}
  \centering
  \begin{tabular}{cccc}
    \hline
    Performance metric                                                    & \textbf{\begin{tabular}[c]{@{}c@{}}Building \\ hit rate {[}\%{]}\end{tabular}} & \begin{tabular}[c]{@{}c@{}}Floor \\ hit rate {[}\%{]}\end{tabular} & 3D error {[}m{]} \\ \hline
    \begin{tabular}[c]{@{}c@{}}RNN with \\ augmented dataset\end{tabular} & \textbf{100}                                                                   & \textbf{94.20}                                                     & \textbf{8.42}    \\
    RNN~\cite{2021hierarchical}                                           & \textbf{100}                                                                   & 95.23                                                              & 8.62             \\
    MOSAIC                                                                & 98.65                                                                          & 93.86                                                              & 11.64            \\
    HFTS                                                                  & \textbf{100}                                                                   & \textbf{96.25}                                                     & 8.49             \\
    ICSL                                                                  & \textbf{100}                                                                   & 86.93                                                              & 7.67             \\ \hline
  \end{tabular}
\end{table}
%%%

\subsection{Offline Evaluation}
\label{sec:offline-evaluation}
\begin{figure*}[htb]
  \begin{minipage}[c]{.33\textwidth}
    \centering
    \includegraphics[width=\textwidth,trim=80 0 80 0,clip=true]{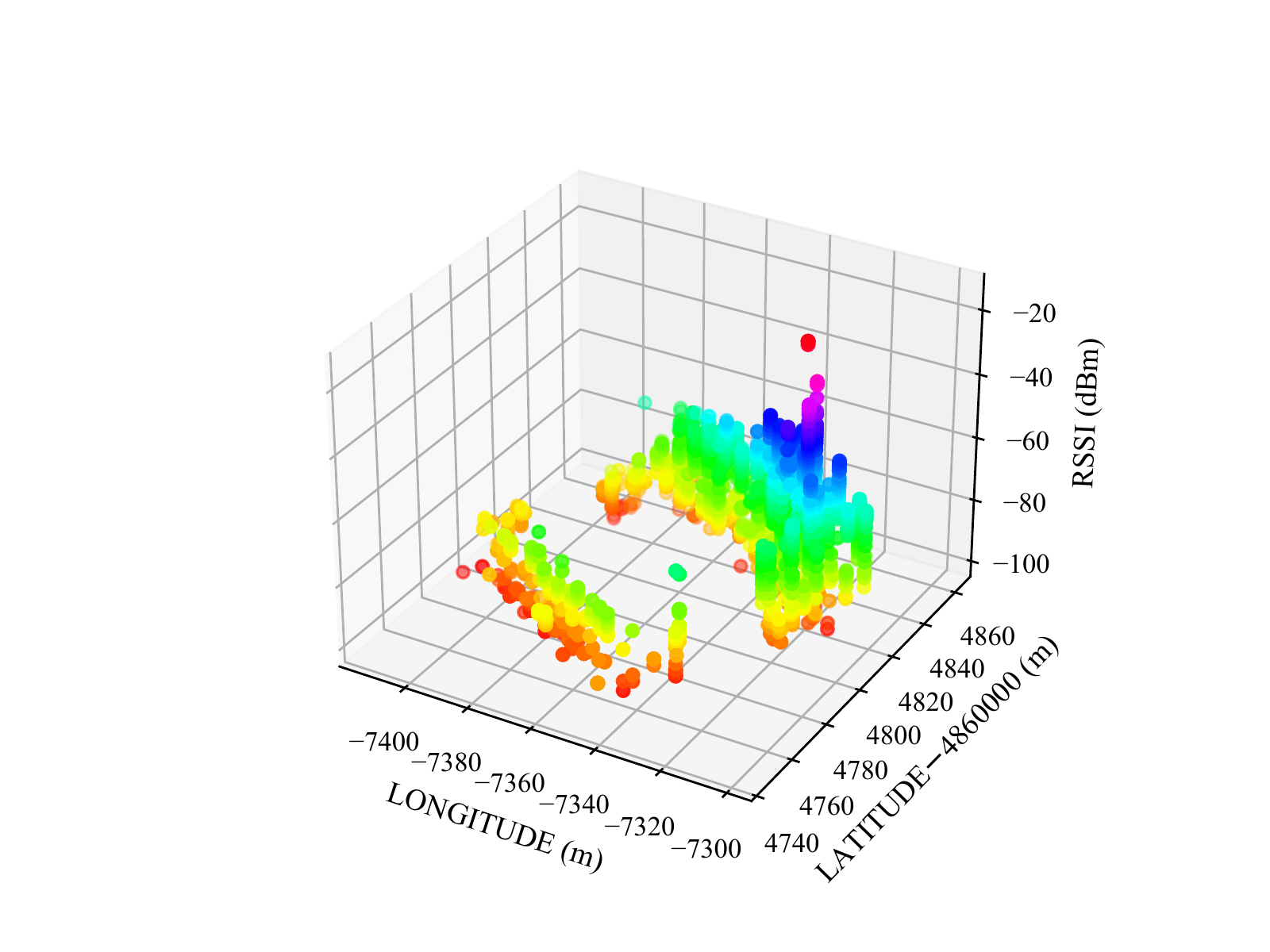}\\
    \scriptsize{(a) Original}
  \end{minipage}
  \begin{minipage}[c]{.33\textwidth}
    \centering
    \includegraphics[width=\textwidth,trim=80 0 80 0,clip=true]{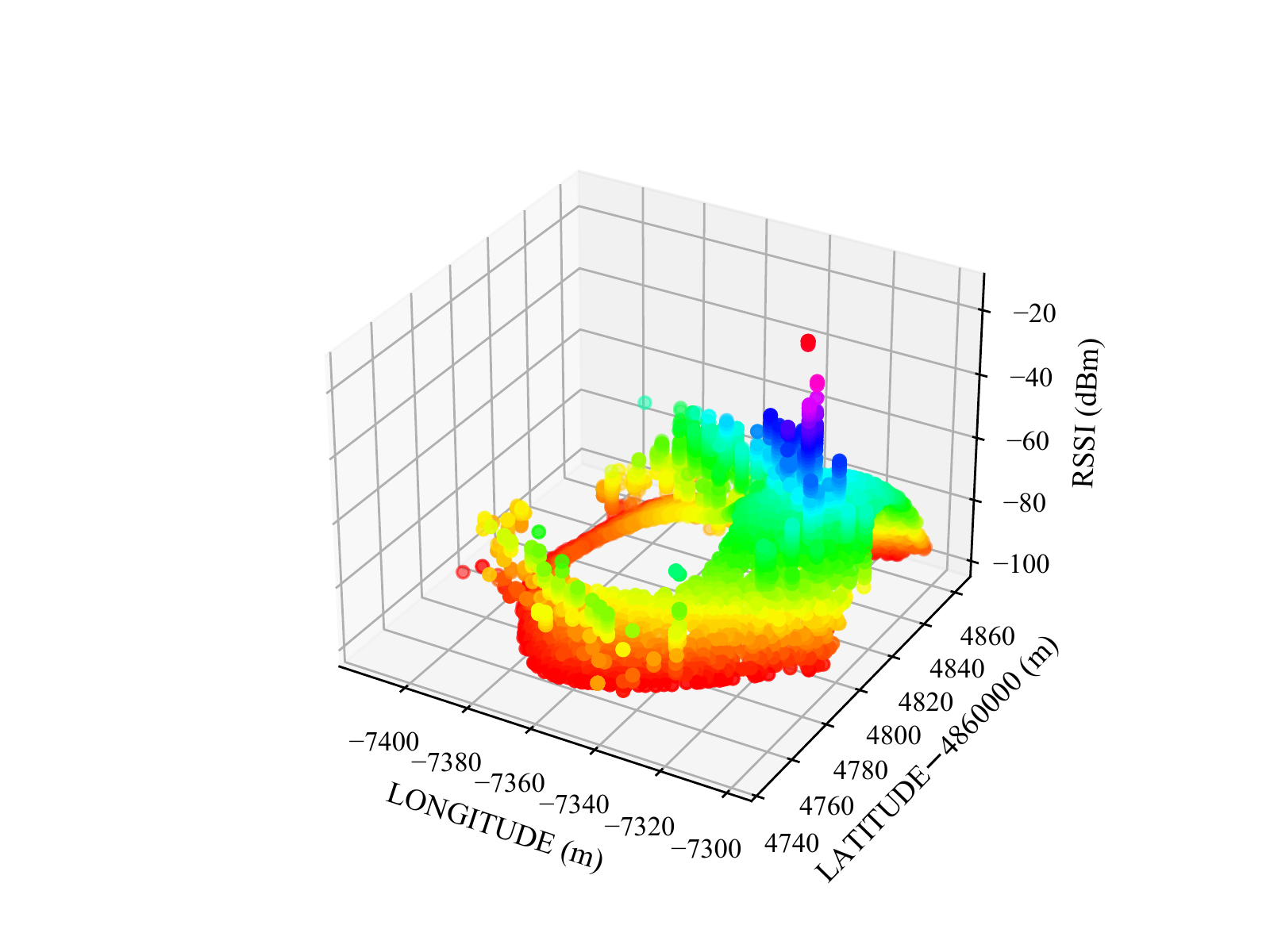}\\
    \scriptsize{(b) RBF}
  \end{minipage}
  \begin{minipage}[c]{.33\textwidth}
    \centering
    \includegraphics[width=\textwidth,trim=80 0 80 0,clip=true]{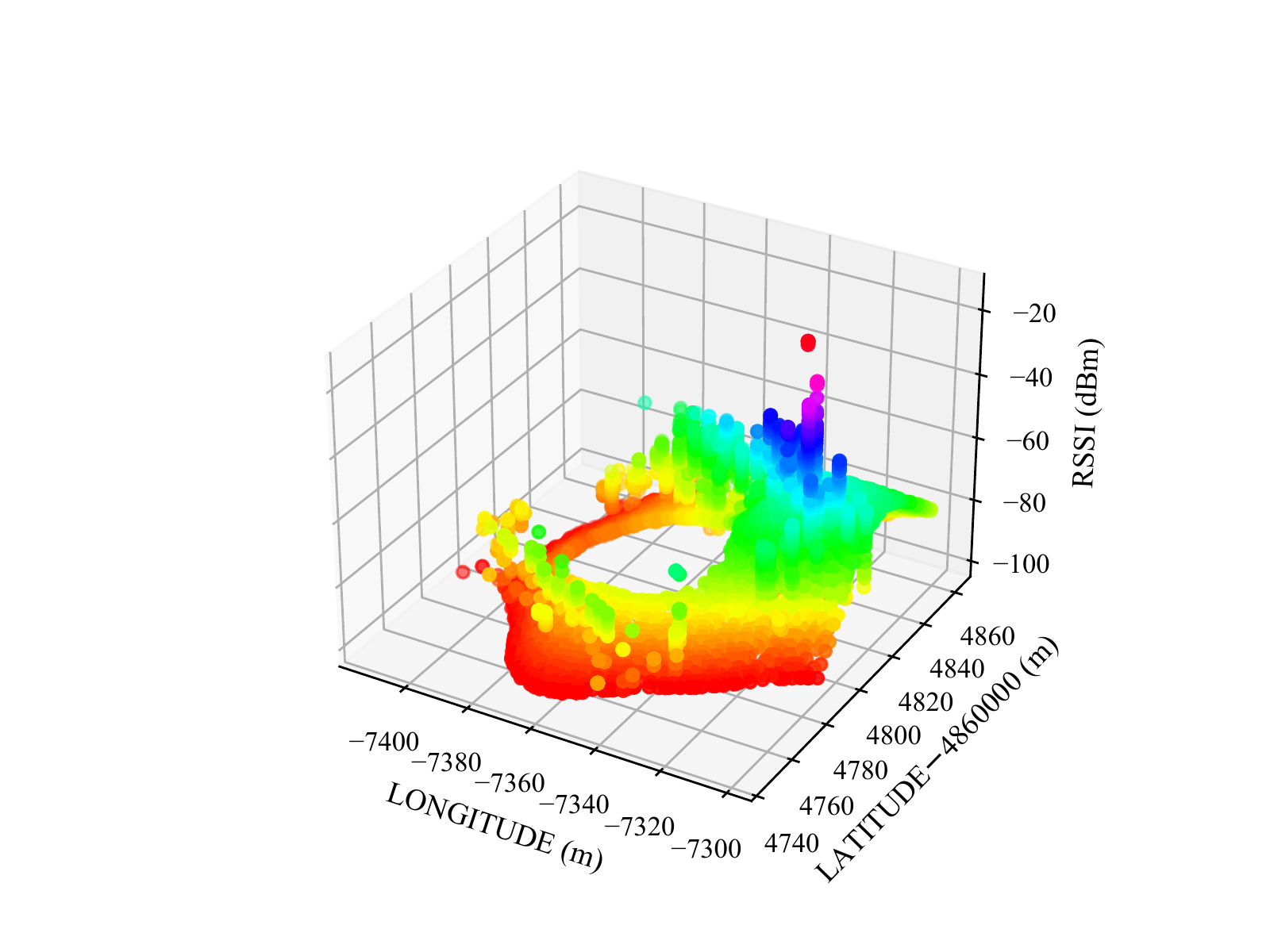}\\
    \scriptsize{(c) Matern32}
  \end{minipage}
  \newline
  \begin{minipage}[c]{.33\textwidth}
    \centering
    \includegraphics[width=\textwidth,trim=80 0 80 0,clip=true]{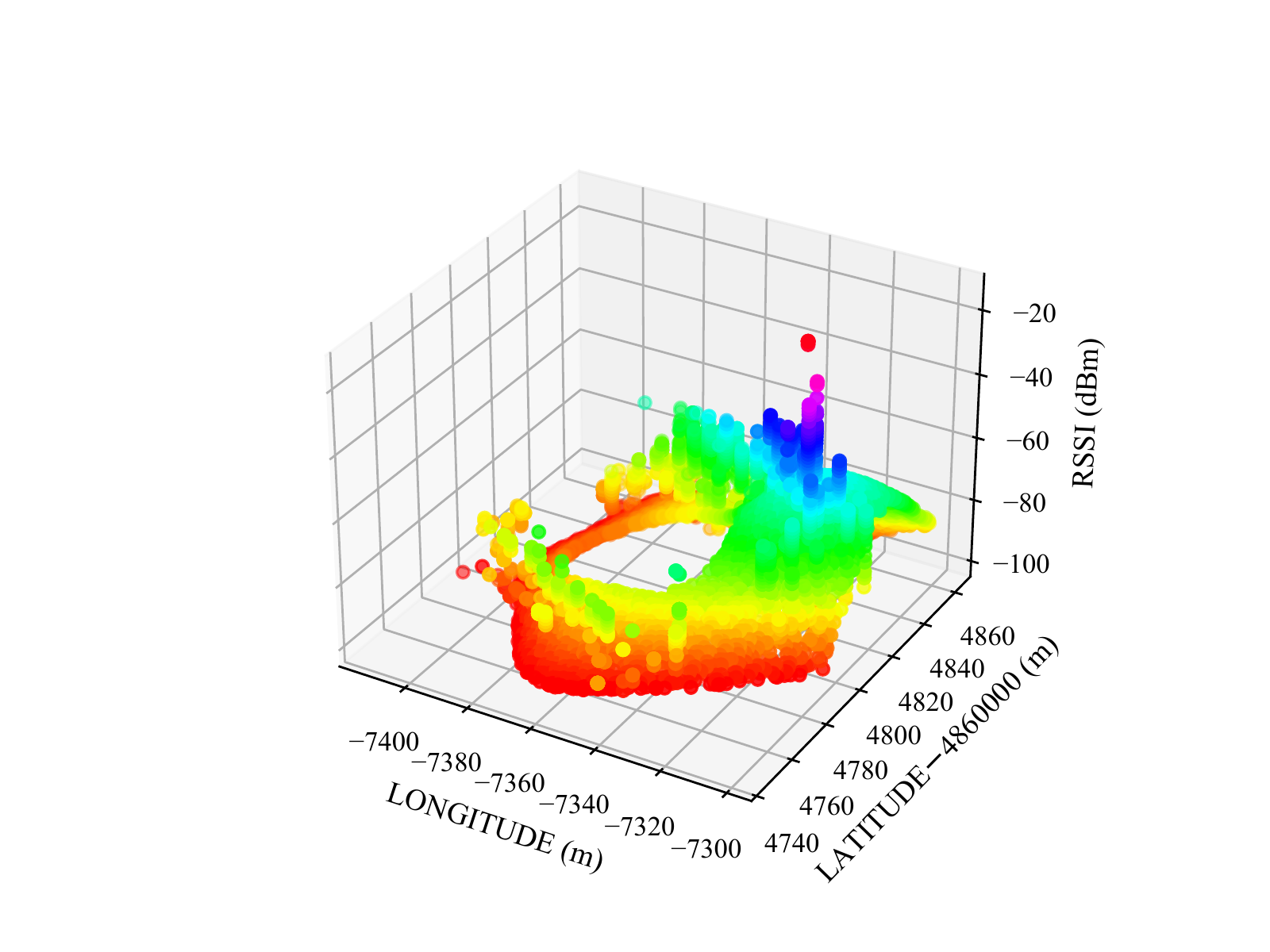}\\
    \scriptsize{(d) Matern52}
  \end{minipage}
  \begin{minipage}[c]{.33\textwidth}
    \centering
    \includegraphics[width=\textwidth,trim=80 0 80 0,clip=true]{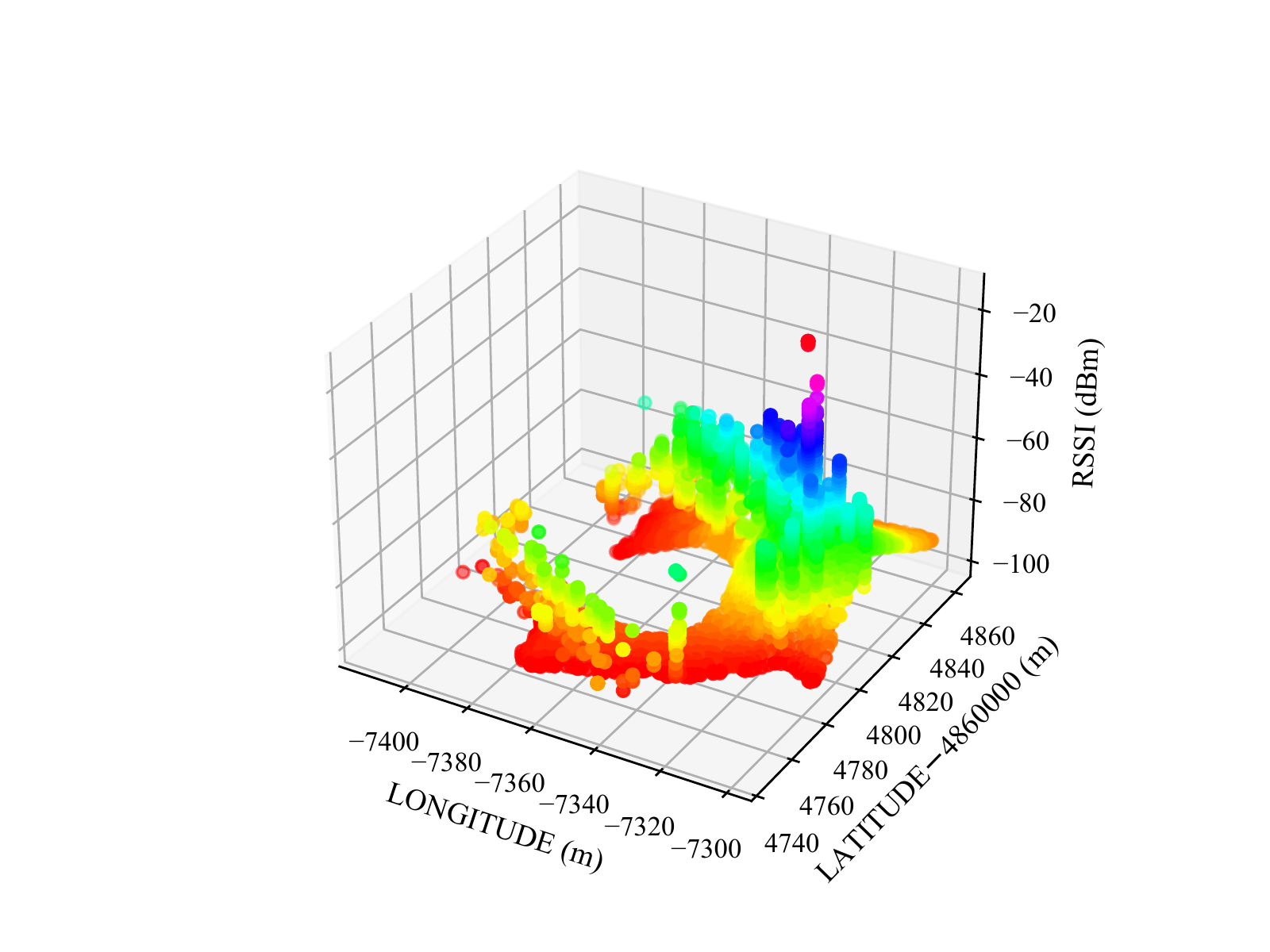}\\
    \scriptsize{(e) OU}
  \end{minipage}
  \begin{minipage}[c]{.33\textwidth}
    \centering
    \includegraphics[width=\textwidth,trim=80 0 80 0,clip=true]{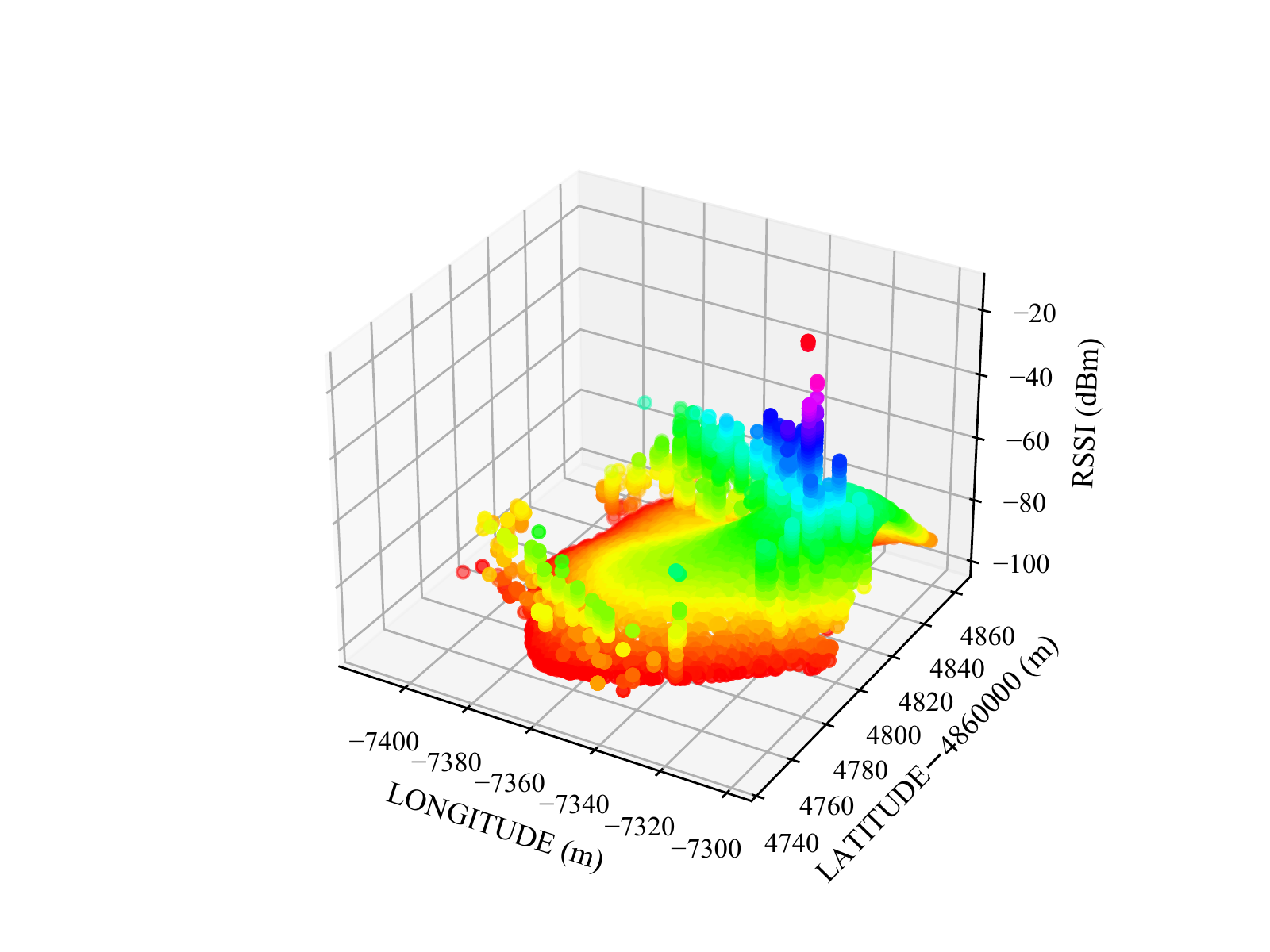}\\
    \scriptsize{(f) RQ}
  \end{minipage}
  \caption{The original RSSI value of WAP500 against Longitude and Latitude and
    augmented value with different kernel functions.}
  \label{fig:wap500rssi}
\end{figure*}

\begin{figure*}[htb]
  \begin{minipage}[c]{.33\textwidth}
    \centering
    \includegraphics[width=\textwidth,trim=80 0 80 0,clip=true]{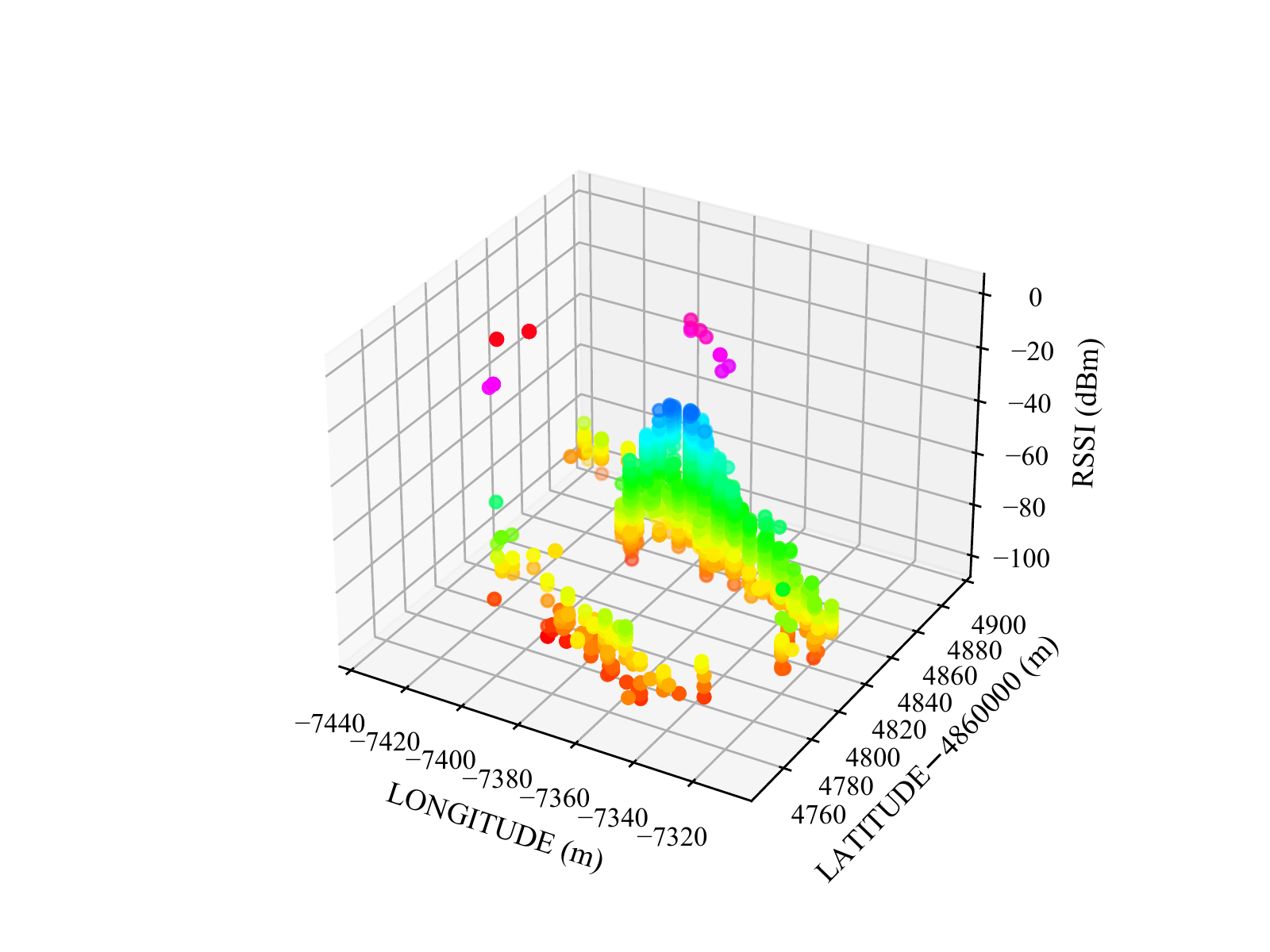}\\
    \scriptsize{(a) Original}
  \end{minipage}
  \begin{minipage}[c]{.33\textwidth}
    \centering
    \includegraphics[width=\textwidth,trim=80 0 80 0,clip=true]{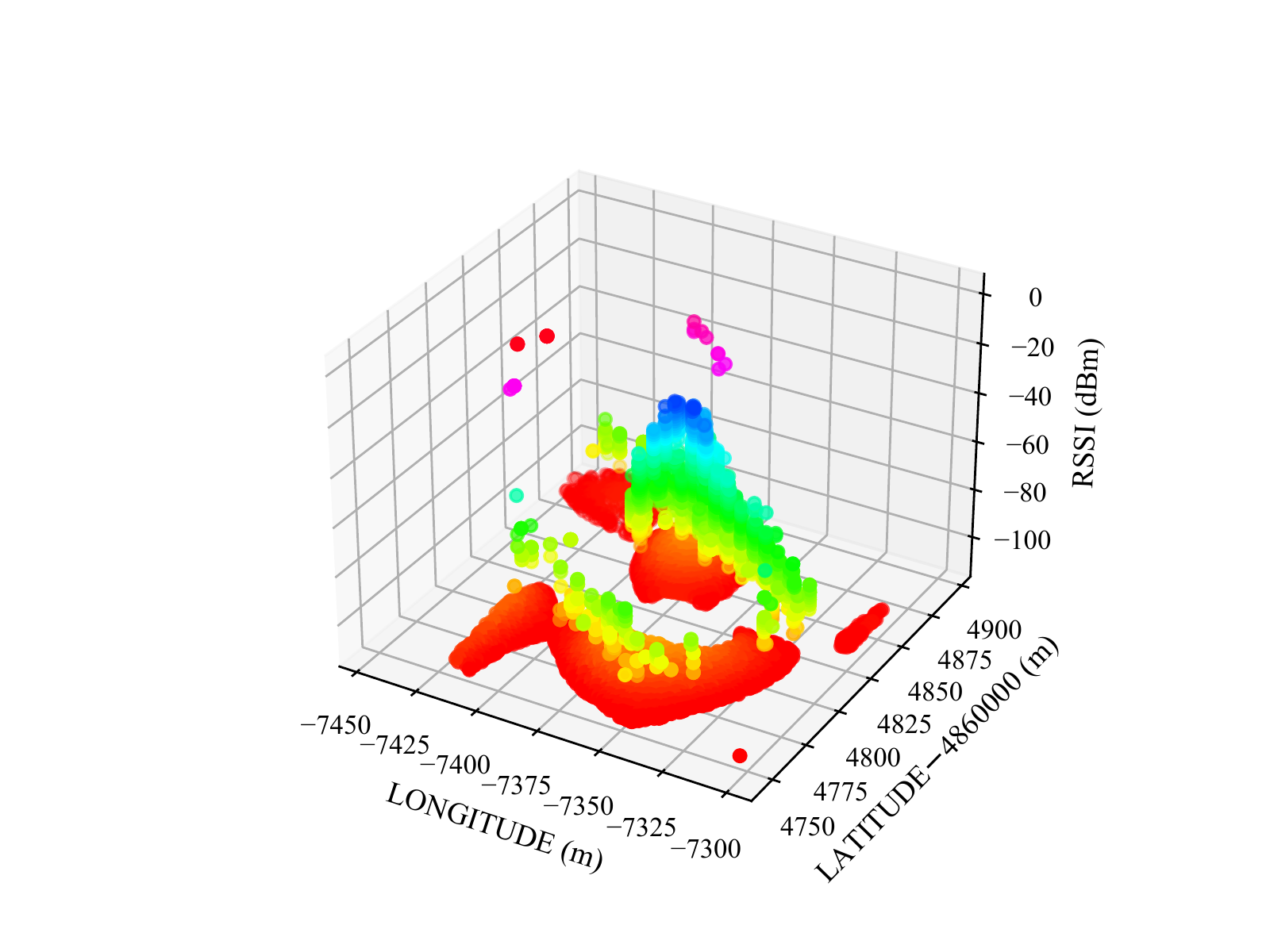}\\
    \scriptsize{(b) RBF}
  \end{minipage}
  \begin{minipage}[c]{.33\textwidth}
    \centering
    \includegraphics[width=\textwidth,trim=80 0 80 0,clip=true]{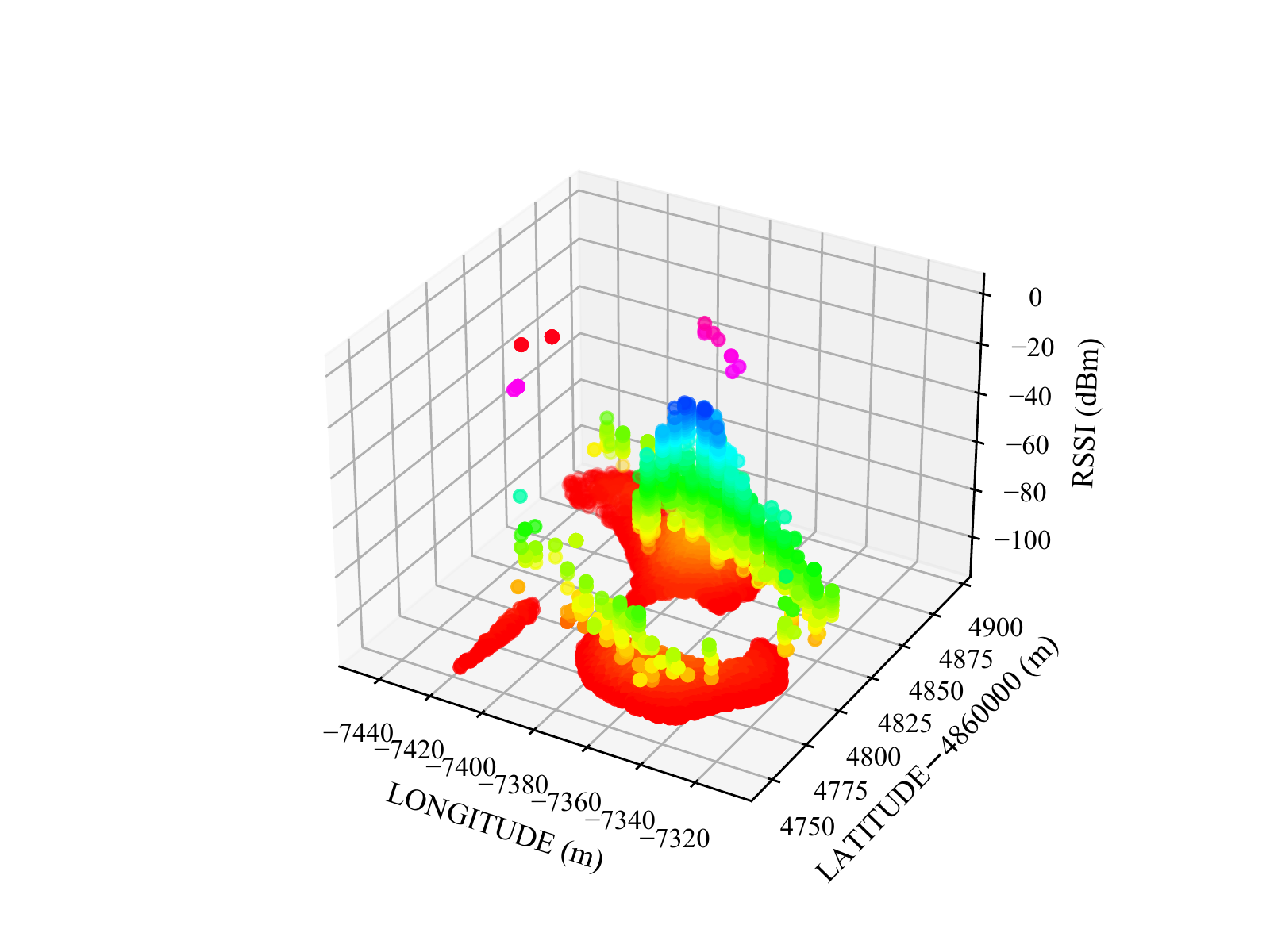}\\
    \scriptsize{(c) Matern32}
  \end{minipage}
  \newline
  \begin{minipage}[c]{.33\textwidth}
    \centering
    \includegraphics[width=\textwidth,trim=80 0 80 0,clip=true]{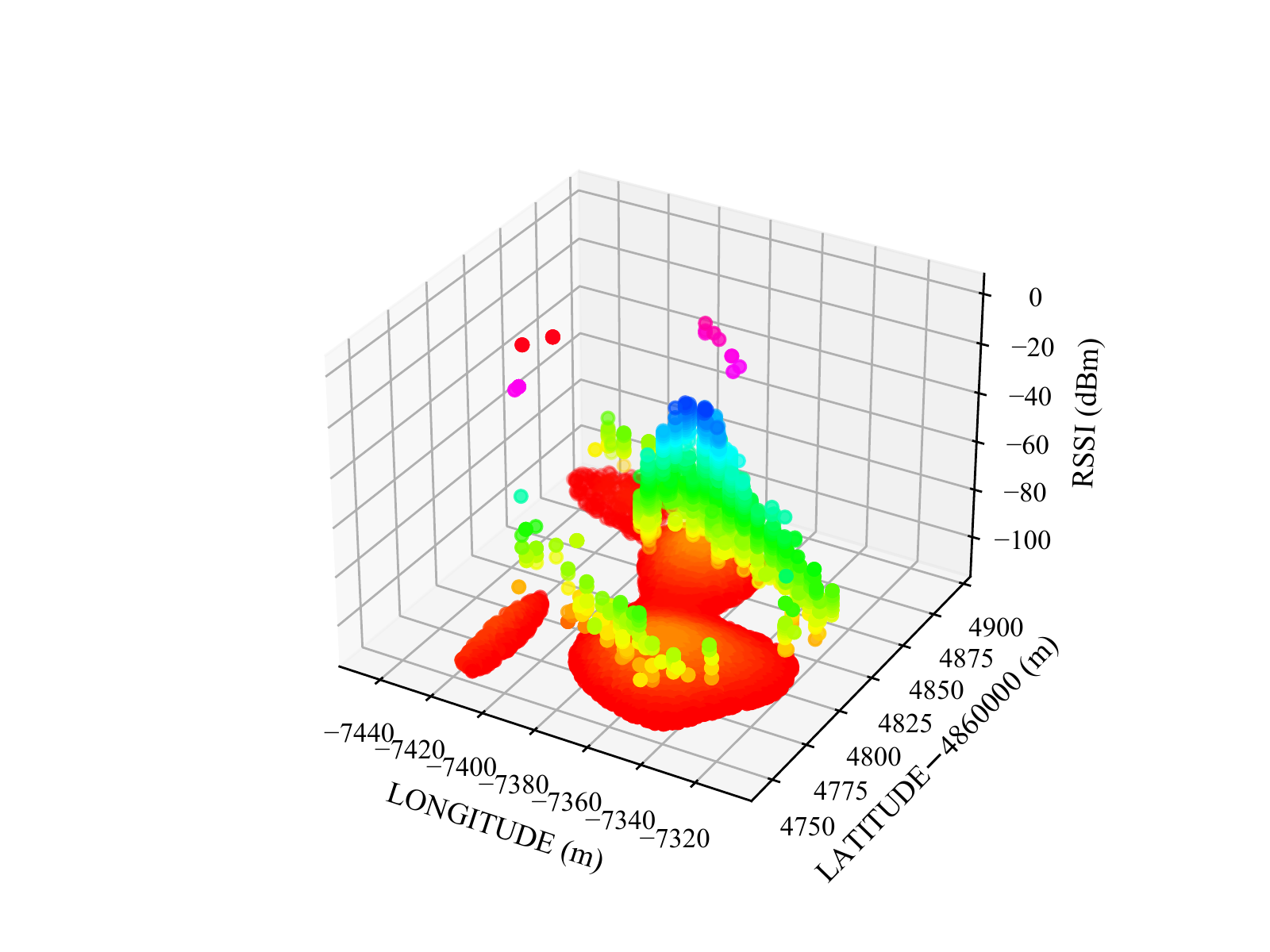}\\
    \scriptsize{(d) Matern52}
  \end{minipage}
  \begin{minipage}[c]{.33\textwidth}
    \centering
    \includegraphics[width=\textwidth,trim=80 0 80 0,clip=true]{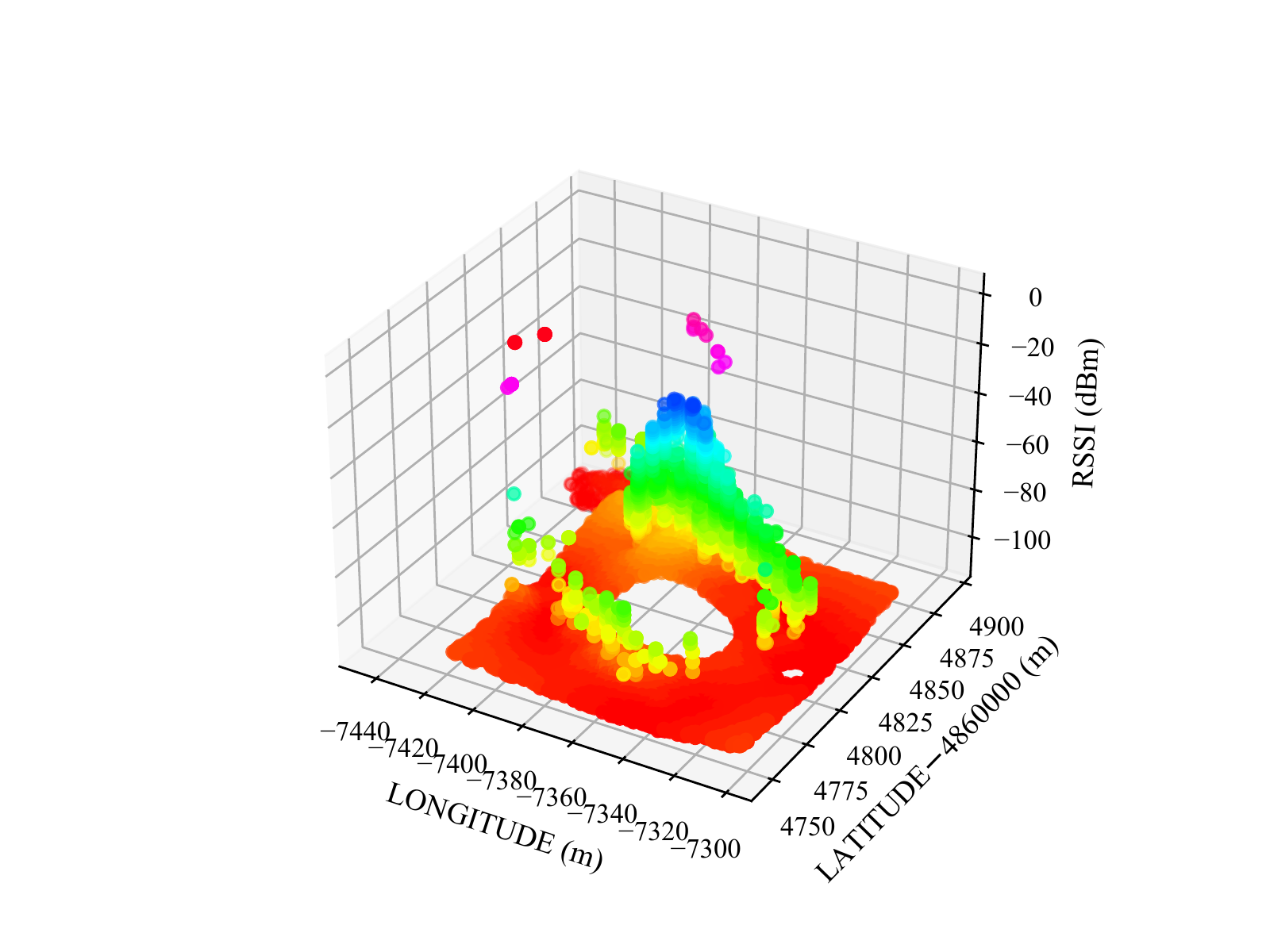}\\
    \scriptsize{(e) OU}
  \end{minipage}
  \begin{minipage}[c]{.33\textwidth}
    \centering
    \includegraphics[width=\textwidth,trim=80 0 80 0,clip=true]{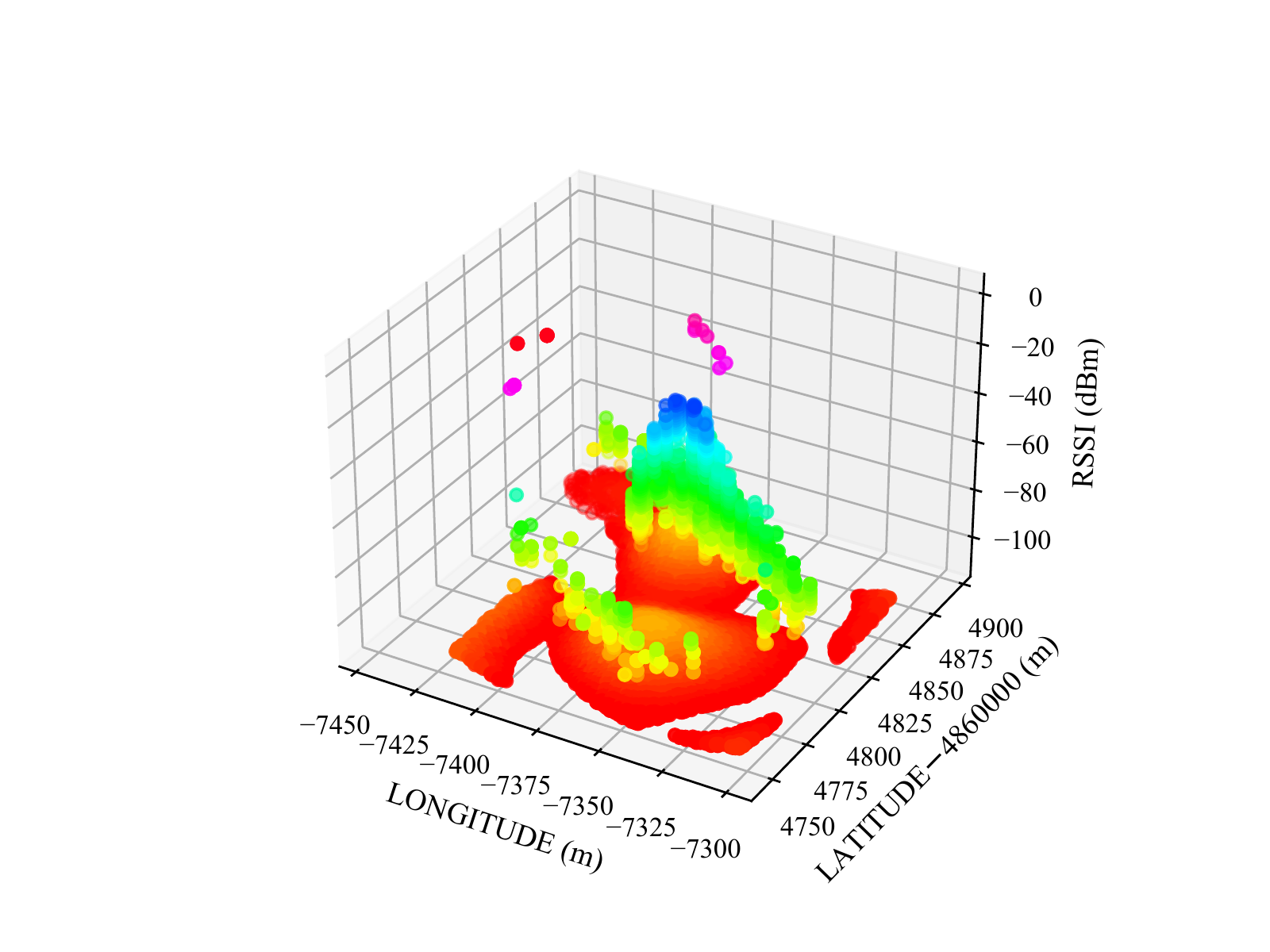}\\
    \scriptsize{(f) RQ}
  \end{minipage}
  \caption{The original RSSI value of WAP11 against Longitude and Latitude and
    augmented value with different kernel functions.}
  \label{fig:wap11rssi}
\end{figure*}

%%%
It is worth noting that since the UJIIndoorLoc database includes a total of 520
WAPs and the data augmentation using kernel functions with similar structures
have limited changes on a single WAP data, usually the variance change of the
data is less than one. Therefore, although we used MOGP to regress all the WAPs,
we did not show all of them but selected some of the WAPs with significant and
representative changes for visualization. The offline evaluation focuses on
comparing the RSSI values of the same WAP under different kernel functions. More
perspectives are mentioned in the online evaluation
Section~\ref{sec:online-evaluation} including kernel functions, hyperparameters,
MOGP models, and the impact of the augmentation ratio on data augmentation. The
principle of selecting WAP is to compare the variance of the augmented data with
that of the original data. The variance of WAP can be defined as,
%%%
\begin{equation}
  \mathbf{Var}(\mathbf{RSSI_{n}}).
  \label{equ:offline_var}
\end{equation}
%%%
Fig.~\ref{fig:wap500rssi} shows the original RSSI value of WAP500 and augmented
data with different kernel functions. Note that in order to highlight changes,
data points with RSSI values less than or equal to -110 are not drawn, which
results in a blank at the bottom of Fig.~\ref{fig:wap500rssi} and does not mean
that the area lacks coverage by RPs. Fig.~\ref{fig:wap500rssi}~(a) is original
data which is distinctly different from Fig.~\ref{fig:wap500rssi}~(b)--(f)
because it is stored in a discrete form in the database.
Fig.~\ref{fig:wap500rssi} and Fig.~\ref{fig:wap11rssi} show similar red bases
since the augmentation results appear as a Gaussian hyperplane, and the
smoothness of the kernel function causes the augmented data to be generated by
gradually fitting the RSSI value of -110~dBm to the target value, which is more
apparent in Fig.~\ref{fig:wap11rssi}. The difference between the several kernel
functions, therefore, lies mainly in their ability to describe the mutant data,
reflected in Fig.~\ref{fig:wap500rssi}, as the sharpness with which the cave
region at the bottom is connected to the surrounding data.
Fig.~\ref{fig:wap500rssi}~(f) does not have such a cavity which can also
correspond to the 3D localization error mentioned in
Table~\ref{tbl:different_kerenls}. Considering a total of 520 WAPs, we therefore
consider Fig.~\ref{fig:wap500rssi}~(b)--(e) having a similar profile to be
insufficient visual evidence of reduced 3D localization error in
Table~\ref{tbl:different_kerenls}.

%%%
%%%

Fig.~\ref{fig:wap11rssi} shows the RSSI values for WAP11 and combined with the
Table~\ref{tbl:different_kerenls} we believe that
Fig.~\ref{fig:wap11rssi}~(e)--(f) reflect that the OU and RQ kernel functions
lack of ability to fit the tail data.

%%%
%%%

\section{Conclusions}
\label{sec:conclusions}
%%%
In this paper, we have proposed multidimensional fingerprint data augmentation
for indoor localization in a large-scale building complex based on MOGP
% to improve the performance of Wi-Fi fingerprint-based indoor localization and
% reduce the time and labor cost of constructing fingerprint databases,
% especially under the current panemic situation of COVID-19.
and systematically investigated the impact of the various aspects of MOGP-based
augmentation on localization accuracy.

Through the extensive experiments using the UJIIndoorLoc database~\cite{UJI}
and the-state-of-the-art neural network indoor localization model based on the
hierarchical RNN~\cite{2021hierarchical}, we first investigated the impact of
MOGP kernel functions and their hyperparameters on the localization performance
and found that Matern52 with the variance of 1 and the length-scale of 10
provides the best performance in the case of a single kernel function. As for
MOGP models, we focused on the impact of the number of kernels $Q$ of LMC (with
ICM being the special case of LMC for $Q{=}1$) and found that the localization
error becomes minimum when $Q$ is equal to the number of MOGP outputs $T$ for
the UJIIndoorLoc database; we also found that $Q{=}2$ can provide decent
localization performance (i.e., second only to $Q{=}T$ in
Table~\ref{tbl:model}) and hit the right balance between performance and
efficiency as suggested
in~\cite{fricker2013multivariate,nguyen2014collaborative}.

The impact of data augmentation ratio was investigated, too, in order to
explore the extent to which we can augment a fingerprint database with
synthetically-generated fingerprints without diluting or losing the statistical
characteristics of real ones. The experimental results suggests that we can
generate synthetic RSSI fingerprint data up to ten times the original
data---i.e., the augmentation ratio of 10---through the proposed
multidimensional MOGP-based data augmentation without significantly affecting
the indoor localization performance compared to that of the original data
alone. The result of this investigation of data augmentation is especially
important because this means that we can extend the spatial coverage of the
combined RPs of a fingerprint database using the proposed MOGP-based data
augmentation and thereby could improve the localization performance at the
locations that are not part of the test dataset.

During our investigation of the impact of various aspects of MOGP-based data
augmentation on localization accuracy, we focused our investigation of MOGP on
the linear models of ICM and LMC and based the experiments only on the
UJIIndoorLoc database. Our investigation in this paper, therefore, could be
extended with other MOGP models and kernel functions better suited for indoor
localization and multi-building and multi-floor databases.

One important issue of the existing fingerprint databases not considered in
this paper is the inadequate consideration of interference factors, which are
often time-varying: In large shopping malls and transport hubs, dense crowds of
moving people are the main interference, while in underground car parks, a
large number of temporary access points are the main interference. Fingerprint
data augmentation taking into account those time-varying interference factors
is another interesting topic for further research.
% And in complex building structures, construction materials can significantly
% weaken the signal.  A time factor also gets less attention; considered to
% reduce human resource costs and time costs, with most of collect fingerpint
% data without recording the RSS average of the measurement localization or
% choosing a multi-fingerprint fusion method to cope with the multimodal
% temporal probability distribution of RSS~\cite{RSS_measure_time}.

\section*{Acknowledgment}
%%%
This work was supported in part by Postgraduate Research Scholarship (under
Grant PGRS1912001) and Key Program Special Fund (under Grant KSF-E-25) of Xi'an
Jiaotong-Liverpool University.

%%% References
% - with BiBTeX
% \bibliographystyle{IEEEtran}%
% \bibliography{IEEEabrv,reference}%


\end{document}